\documentclass[
reprint,
superscriptaddress,
amsmath,amssymb,
aps,
prb,
floatfix,
groupedaddress
]{revtex4-2}

\usepackage{color}
\usepackage{graphicx}
\usepackage{dcolumn}
\usepackage[normalem]{ulem}
\usepackage{bm}
\usepackage{mathtools}
\usepackage{bbold}
\usepackage{enumerate}
\usepackage{multirow}
\usepackage{subfigure}
\usepackage{hyperref}

\begin{document}

\preprint{}

\title{The Low-Temperature Phenomenology of Gap Inhomogeneity in the Cuprate Superconductors: High-Energy Granularity, Low-Energy Homogeneity, and Spectral Kinks}

\author{Miguel Antonio Sulangi}
\affiliation{
	National Institute of Physics, University of the Philippines Diliman, Quezon City 1101, Philippines
}

\date{\today}

\begin{abstract}
 
Scanning tunneling spectroscopy experiments on a number of cuprate superconductors have revealed that these materials are highly inhomogeneous. However, even though this inhomogeneity is well-characterized experimentally, a theoretical understanding of the effect of an inhomogeneous superconducting $d$-wave order parameter on various observables is still not complete. Here, we focus on the particular role played by the length scale of superconducting order-parameter inhomogeneity. We make use of a model involving square patches tiling the system, with each patch hosting a broadly distributed random value of the $d$-wave parameter. By using large-scale simulations, we are able to study how the size of the patches affects the correspondence between various measures of the superconducting gap and the underlying order parameter. If the length scale of the inhomogeneity is smaller than the average superconducting coherence length, the resulting $d$-wave superconductor is homogeneous. However, when the order parameter varies on the scale of the coherence length, we find the emergence of a striking low-energy--high-energy dichotomy, in which the low-energy regime is homogeneous while the high-energy states are strongly inhomogeneous. Kinks in the local spectra are found at the energy demarcating the homogeneous-inhomogeneous transition. We also observe that the gap extracted from the low-energy slope of the local density of states is extremely uniform. We find in both of these regimes that the distribution of the spectral gap is narrower than that of the order parameter; these start to match only when the size of the patches becomes parametrically larger than the coherence length. We comment on the applicability of these results to the cuprates, discuss the limitations of the inhomogeneous $d$-wave model, and point out where beyond-mean-field correlation effects are likely to be present in addition to inhomogeneity.

\end{abstract}

\maketitle

\section{\label{sec:level1}Introduction}

Scanning tunneling spectroscopy (STS) has been a hugely invaluable tool for uncovering many of the mysteries held by the high-$T_c$ cuprates \cite{schmidt2011electronic,fujita2011spectroscopic}. The real-space resolution afforded by state-of-the-art STS measurements has facilitated numerous important discoveries about these materials, such as the presence of strong inhomogeneity in the spectral gap \cite{pan2001microscopic,howald2001inherent,lang2002imaging,mcelroy2005coincidence,mcelroy2005atomic,fang2006gap,gomes2007visualizing,alldredge2008evolution,pasupathy2008electronic,pushp2009extending,parker2010nanoscale,alldredge2013universal,li2022strongly,tromp2023puddle}, broken-symmetry states coexisting with the superconducting state \cite{hoffman2002four,howald2003coexistence,howald2003periodic,hamidian2016detection,edkins2019magnetic,du2020imaging}, and quasiparticle scattering interference (QPI) \cite{hoffman2002imaging,wang2003quasiparticle,capriotti2003wave,mcelroy2003relating,zhu2004power,kohsaka2008cooper,fujita2014simultaneous,he2014fermi,webb2019density,ye2024emergent}. The maturation of this technique has allowed its use beyond the low-temperature regime of the cuprates where it originally found success: successful applications of the technique to both high temperatures \cite{gomes2007visualizing,pasupathy2008electronic,pushp2009extending,parker2010nanoscale} and heavy overdoping \cite{li2022strongly,tromp2023puddle,ye2024emergent} have uncovered strong inhomogeneity even in these regimes. The story emerging from three decades of STS work on the cuprates is one in which a number of seemingly conflicting ingredients are present in some form deep inside the superconducting state: patchy inhomogeneity (once likened to ``quantum salad dressing'' \cite{zaanen2002quantum} due to the appearance of what appears to be electronic phase separation), density-wave order, and, somehow, a sufficient amount of homogeneity needed to allow for the existence of QPI.  How these three apparently mutually incompatible ingredients coexist in the cuprates has been a persistent mystery since their discovery.

Low-temperature differential conductance ($dI/dV$) measurements in the superconducting cuprates generally find a number of features that are present at a wide range of dopings and even in different families of cuprates: 1) the spectral gap is patchy, with a characteristic length scale of several lattice constants \cite{pan2001microscopic,howald2001inherent,lang2002imaging,mcelroy2005coincidence,fang2006gap,gomes2007visualizing,alldredge2008evolution,pasupathy2008electronic,pushp2009extending,parker2010nanoscale,alldredge2013universal,li2022strongly,tromp2023puddle,ye2024emergent}; 2) while the spectra have the V-shaped form characteristic of $d$-wave pairing at the lowest energies, these eventually display ``kinks'' for large-gap regions \cite{howald2001inherent,mcelroy2005coincidence,fang2006gap,alldredge2008evolution,pushp2009extending,alldredge2013universal,ye2024emergent}; 3) there is a negative correlation between the spectral gap and the height of the coherence peak, so that small-gap regions display comparatively tall and sharp peaks and large-gap regions have stubbier ones \cite{lang2002imaging,mcelroy2005coincidence,fang2006gap,alldredge2008evolution,ye2024emergent}; 4) the width of the spectral-gap distribution generally scales with the mean spectral gap, so the distribution is narrow at overdoping (where the average gap is small) and wide at underdoping \cite{mcelroy2005coincidence,alldredge2008evolution,gomes2007visualizing,pasupathy2008electronic,pushp2009extending}; and, perhaps most important, 5) the spectra are highly homogeneous at low bias voltage, with inhomogeneity appearing only at a higher energy scale that is set roughly by the location of the kink but is also parametrically smaller than the average spectral gap \cite{mcelroy2005coincidence,kohsaka2008cooper,alldredge2008evolution,pushp2009extending,alldredge2013universal}. The last observation dovetails nicely with the presence of quasiparticle scattering interference at low energies which is eventually lost at an energy that matches the scale where homogeneity is lost \cite{kohsaka2008cooper,fujita2014simultaneous,he2014fermi,webb2019density}. These observations are qualitatively consistent with the presence of \emph{homogeneous} $d$-wave superconductivity at low energies, which eventually gives way to an \emph{inhomogeneous} regime at high energies characterized by broken symmetry \cite{mukhopadhyay2019evidence}.

This dichotomy between the low- and high-energy states in the underdoped cuprates has been interpreted as that arising from two distinct types of excitations. In this scheme, the low-energy states are simply the Bogoliubov quasiparticles of a homogeneous $d$-wave superconductor, while the high-energy states correspond to mysterious ``pseudogap'' states which break translational invariance in real space and which ostensibly originate from the parent Mott insulating state \cite{kohsaka2008cooper,fujita2014simultaneous,he2014fermi,webb2019density}. This explanation is a particularly compelling one since this dichotomy (in \emph{energy}) is structurally similar to the dichotomy one sees in \emph{temperature} in the underdoped regime: the low-temperature state is a $d$-wave superconductor, while the high-temperature state (the pseudogap) hosts translational-symmetry-breaking charge order and features the absence of any phase coherence. Consequently, the high-energy states imaged in these $dI/dV$ measurements are manifestations at low temperature of the same states that are present in the high-temperature regime above $T_c$. This is also consistent with the picture emerging from angle-resolved photoemission spectroscopy (ARPES), which finds evidence for two distinct gaps---one corresponding to the superconductor, and another corresponding to the pseudogap \cite{tanaka2006distinct,lee2007abrupt}.

As compelling as this picture is, one is inevitably confronted with the question of how homogeneous $d$-wave superconductivity can be made to coexist with an exotic, spatially fluctuating pseudogap state at the lowest temperatures in the first place. This picture would appear to require that these two states be separated in energy, with something akin to a ``mobility edge'' separating the delocalized (``$\mathbf{k}$-space'') low-energy states from the localized (``$\mathbf{r}$-space'') high-energy states. Given that the cuprates are strongly correlated materials, it is not hard to imagine that such a possibility can arise from a very complicated microscopic model that can generate both the low-energy translationally-invariant superconductor and the high-energy translational-symmetry-breaking pseudogap simultaneously. The fact that simple paradigmatic models such as the Hubbard model do not yield such behavior easily can lead one to think that any putative model explanation for this necessarily requires a complicated mechanism underlying this separation of scales in energy. (We must mention that an important advance is the work by Lee et al., which makes use of a model of ``glassy nematicity'' and, within a mean-field framework where the nematic order is disordered but the pairing interaction is not, is able to reproduce to some extent some of the features of this low-energy--high-energy dichotomy seen in STS spectra taken from the cuprates \cite{lee2016cold}. However, in their work, they find the homogeneous regime to lie within all energies less than the average spectral gap and the inhomogeneous regime to reside outside the spectral gap, which is a feature that is \emph{not} seen in the cuprates---in experiment, the homogeneous-inhomogeneous transition occurs at a characteristic energy scale that is \emph{smaller} than the average spectral gap \cite{alldredge2008evolution,alldredge2013universal}.)

In this paper, we take a different tack: we leave the question of modeling the exotic, symmetry-broken pseudogap state aside and instead study a much simpler mean-field model of $d$-wave superconductivity with strong order-parameter inhomogeneity, but without any chemical-potential or hopping disorder or coexisting order. We wish to study the extent to which the aforementioned STS phenomenology can be captured within mean-field theory by a \emph{minimal model} consisting only of strong order-parameter inhomogeneity, and we pay close attention to the role played by the \emph{length scale} of the inhomogeneity. We do not make any assumptions about the origin of the inhomogeneity. Instead, we make use of an exceedingly simple model of gap inhomogeneity: we divide the system into equally shaped square patches, and to each patch we assign a random value of the $d$-wave order parameter, taken from a uniform distribution. (This is similar in spirit to the patch model used by Fang et al. \cite{fang2006gap} and by Alvarez and Dagotto \cite{alvarez2008fermi}. It should be said that another self-consistent, \emph{ex post facto} justification for the use of this model is to see how a most artificial-looking inhomogeneity model can in fact reproduce experimental spectra without much fine-tuning, as can be seen later.) The distribution of order-parameter values is fixed such that these range from zero to twice the mean; the only parameter being varied is the linear extent $l$ of the patch. The main utility of the square-patch model is the ease with which one can tune the length scale of the inhomogeneity, as opposed to other models such as the Yukawa-like model used by Nunner et al. \cite{nunner2005dopant,nunner2006fourier,andersen2006thermodynamic}. We use Green's function-based techniques to efficiently obtain the local density of states (LDOS) for very large systems (with around $3\times 10^5$ sites), allowing us to calculate spectral observables and statistical correlations among these observables with much accuracy. The observables we focus on are the LDOS spectra themselves and convenient experimental proxies for the true superconducting order parameter: the spectral gap, which is the location of the coherence peak, and the low-energy gap, which is the value of the gap one can extract from the low-energy slope of the LDOS, which in a bulk $d$-wave superconductor is inversely proportional to the order parameter. 

We find three generic regimes, depending on how large the patch length $l$ is compared to the \emph{average} superconducting coherence length $\xi$. If $l$ is less than $\xi$, the quasiparticle response of the superconductor to this order parameter profile is remarkably spatially uniform. In this regime, the LDOS from site to site does not vary appreciably. The order-parameter proxy observables show only very weak variations, and the distributions of both of these quantities are very narrow compared with that of the underlying $d$-wave order parameter. The spectra are effectively averaged over a coherence volume $\xi^2$, so the underlying atomic-scale variations in the order parameter are smoothed over in the quasiparticle response. If, on the other hand, $l$ is around the same magnitude as $\xi$, strikingly different behavior is observed which matches that seen in the superconducting cuprates. The LDOS at low energies is very uniform throughout the system, and deviations from uniformity only arise after a threshold energy is exceeded, beyond which the LDOS becomes strongly spatially fluctuating. We find that the kink energy demarcating the homogeneous-inhomogeneous crossover is generally set by the smallest set of gaps in the system, something that is also seen in experiment. In this regime, the statistical correlation between the local spectral gap and the height of the coherence peak is negative, similar to what is seen in STS experiments. Finally, in the regime where $l$ is larger than $\xi$, the behavior of the LDOS at a site where the local $d$-wave order parameter has a value $\Delta$ gradually reverts to that of a \emph{bulk} $d$-wave superconductor with the same value of the order parameter. In this regime, the statistical correlations among the various observables also approach that expected from a bulk superconductor, with the proxy observables becoming very closely matched to the true order parameter. The locations of the kinks get pushed down all the way to zero energy with increasing $l$.

We further find that the system averages of the spectral gap, the low-energy gap, and the kink energy represent three distinct emergent energy scales which separate from one another as the inhomogeneity length scale is increased. We find that while these three energy scales all coincide with one another when $l < \xi$, they separate as $l$ is increased. In particular, we see that the average spectral gap increases as $l$ increases up until $l \approx \xi$ is reached, at which point it decreases; that the average low-energy gap decreases with increasing $l$, but generally does not stray far from the mean of the underlying order parameter; and that the average kink energy decreases monotonically with increasing $l$. We also show that the kink energy is generally tied to the appearance of resonances in the LDOS at the smallest-gap regions, which destroy the low-energy homogeneous state.

We discuss in detail some ramifications of our results to the interpretation of STS experiments.  In particular, we argue that in the cuprate superconductors, the behavior of the LDOS spectra and the spectral-gap distributions throughout much of the doping range are consistent with order-parameter patch sizes being in the $l \approx \xi$ regime. As a consequence, both the low-energy homogeneous $d$-wave state and the high-energy inhomogeneous state (but \emph{not} the broken-symmetry density-wave and nematic states) can be explained within the same mean-field framework. We then reexamine the standard interpretation of the spectral gap as being a direct measure of the true superconducting order parameter and point out that in the $\l \approx \xi$ regime relevant to the cuprates, the width of the spectral-gap distribution vastly underestimates the width of the distribution of the underlying $d$-wave order parameter. It is then entirely plausible even that deep within the superconducting state of the cuprates, there can exist patches where the order parameter is very small or even zero, but proximity coupling of these with the surrounding large-order-parameter regions results in a low-energy homogeneous state within these regions that exhibits a finite spectral gap. We also consider estimates of the so-called ``nodal gap'' \cite{pushp2009extending} and find that much of the observed ``two-gap'' phenomenology at low temperatures can be explained even within the inhomogeneous $d$-wave model to a surprisingly decent extent, with the dichotomy between the low-energy and high-energy states playing a crucial role. In addition, we discuss what our results mean for estimates of the local inelastic scattering rate in underdoped cuprates from STS data \cite{alldredge2008evolution,alldredge2013universal}; we find that these estimates are likely an overestimate of the true inelastic scattering rate as these do not invoke the inhomogeneity of the pairing in the estimation process. Finally, we comment on the relation of order-parameter inhomogeneity to the hitherto unsolved phenomenon of QPI extinction \cite{kohsaka2008cooper,fujita2014simultaneous,he2014fermi} and suggest that inhomogeneity is of fundamental importance to this phenomenon in the first place, although we postpone a thorough discussion of this to a future paper.

\section{Model and Methods} \label{model}

\subsection{Mean-Field Hamiltonian and Green's Functions}

Because we intend to study the LDOS and the various quantities one can extract from it, it is imperative that the system sizes in our simulations be as large as possible in order to mitigate any finite-size effects, to mimic the large field of view that state-of-the-art STS experiments can now access, and to obtain robust statistical correlations that can only come from having many (on the order of $10^6$) individual real-space spectra. Our point of departure is a mean-field Bogoliubov-de Gennes (BdG) Hamiltonian describing a $d$-wave superconductor on a square-lattice system of size $N_x \times N_y$:
\begin{equation}
	H = \sum_{\langle ij \rangle} \Big[-\sum_{\sigma} t_{ij}c_{i\sigma}^{\dagger}c_{j\sigma} +\Delta_{ij}^{\ast}c_{i \uparrow}c_{j \downarrow} + \text{h.c.}\Big].
	\label{eq:hamiltonian}
\end{equation}
$t_{ij}$ and $\Delta_{ij}$ are the hopping matrix element and the bond superconducting order parameter, respectively, between sites $i$ and $j$. The notation $\langle ij \rangle$ in the sum denotes a restriction to only pairs of sites $i$ and $j$ that are nearest-neighbors, next-nearest neighbors, or the same as one another. For $t_{ij}$, we disregard higher-order hoppings, so that the normal-state dispersion is $\epsilon_{\mathbf{k}} = -2t(\cos k_x + \cos k_y) - 4t'\cos k_x \cos k_y - \mu$, where $t$, $t'$, and $\mu$ are the nearest-neighbor hopping, next-nearest-neighbor hopping, and the on-site chemical potential, respectively. We set these hopping parameters to be $t = 1$, $t' = -0.3$, and $\mu = -1.03$; the resulting hole doping is $16\%$, in line with the real-world hole doping at which the cuprates generally attain the highest $T_c$. Meanwhile, for $\Delta_{ij}$, we consider only nearest-neighbor pairs of sites. For a \emph{clean} $d$-wave superconductor, $\Delta_{\mathbf{r},\mathbf{r}\pm\hat{x}} = \Delta_0$ and $\Delta_{\mathbf{r},\mathbf{r}\pm\hat{y}} = -\Delta_0$ so that the order parameter in $\mathbf{k}$-space is given by $\Delta_{\mathbf{k}} = 2\Delta_0(\cos k_x - \cos k_y)$. We consider in this paper the more general case where $\Delta_{ij}$ is spatially disordered.

We emphasize once more that we keep the hoppings and the on-site chemical potential constant throughout the entire system, so that the standard assumptions of ``dirty $d$-wave'' theory (where disorder is present through spatial fluctuations of the on-site chemical potential) do not apply here. Instead, inhomogeneity will enter \emph{only} through the ``off-diagonal'' order-parameter matrix elements $\Delta_{ij}$. It is worth mentioning that if we had chemical-potential or hopping disorder, it will naturally generate off-diagonal disorder through the self-consistent solution of the BdG Hamiltonian \cite{atkinson2000gap,atkinson2000details,lee2016cold,sulangi2021correlations,li2021superconductor,pal2023simulating}. However, it is generally found that with weak-to-intermediate levels of chemical-potential disorder (necessary to preserve the observed homogeneity of the superconducting state), the resulting order-parameter inhomogeneity is far too narrowly distributed to be consistent with experimentally obtained gap maps \cite{sulangi2021correlations}. This points to the necessity of including intrinsic disorder in the off-diagonal order-parameter channel that is independent of the variation produced by impurities via self-consistency. On the other hand, a model of glassy nematicity (but with a spatially constant pairing interaction) is able to produce order-parameter variations that look similar to what is seen in experiment \cite{lee2016cold}. In this work, we do not self-consistently determine the order parameter; instead, we impose a fixed order-parameter profile which is characterized by a particular disorder length scale and strength. A fully self-consistent treatment study of pairing-interaction inhomogeneity and its effects on the superfluid density, the order parameter, and the LDOS at various temperatures will be the subject of a forthcoming paper \cite{sulangifuture1}.

We use Green's-function-based techniques to obtain the LDOS in a very fast manner. The Green's function $G(\mathbf{r}_i,\sigma_i | \mathbf{r}_j,\sigma_j | \omega)$ ($\sigma_i$ and $\sigma_j$ are pseudospin indices which for labeling purposes we assume can take on the values $1$ or $2$) in matrix form is 
\begin{equation}
	\mathbf{G} = \Bigl[(\omega + i\Gamma)\mathbf{1} - \mathbf{H}\Bigr]^{-1},
	\label{eq:greensfunction}
\end{equation}
where $\mathbf{H}$ is the Hamiltonian in Eq.~\ref{eq:hamiltonian} written in manifestly Nambu-space form, $\mathbf{1}$ is the identity matrix, $\omega$ is the frequency, $\Gamma$ is the broadening parameter, and a \emph{matrix inverse} is performed. $\mathbf{G}$, $\mathbf{H}$, and $\mathbf{1}$ are all square matrices with dimension $2 N_x N_y \times 2 N_x N_y$. If we impose periodic boundary conditions along the $y$-direction and open boundary condition along the $x$-direction, $\mathbf{H}$ and $\mathbf{G}^{-1}$ can be written as block-tridiagonal matrices. $\mathbf{H}$ in particular can be written in the following suggestive form:
\begin{equation}
\mathbf{H} = \begin{bmatrix} 
	 \mathbf{M}_1 &  \mathbf{T}_1 & \mathbf{0} & \mathbf{0} & \dots & \dots & \mathbf{0} \\
	 \mathbf{T}_1^{\dagger} & \mathbf{M}_2 & \mathbf{T}_2  & \mathbf{0} & \dots & \dots & \mathbf{0} \\
	 \mathbf{0} & \mathbf{T}_2^{\dagger}& \mathbf{M}_3  & \mathbf{T}_3 & \dots & \dots  & \mathbf{0} \\
	 \mathbf{0} & \mathbf{0} & \mathbf{T}_3^{\dagger} & \mathbf{M}_4 & \dots & \dots & \mathbf{0} \\
	 \vdots &\vdots & \vdots & \vdots  & \ddots &  & \vdots \\
	 \vdots &\vdots & \vdots & \vdots & \mathbf{M}_{Nx-2} & \mathbf{T}_{Nx-2} & \mathbf{0} \\
	 \mathbf{0} & \mathbf{0} & \mathbf{0} & \mathbf{0} &  \mathbf{T}_{Nx-2}^{\dagger} & \mathbf{M}_{Nx-1} &  \mathbf{T}_{Nx-1}\\
	 \mathbf{0} & \mathbf{0} & \mathbf{0} & \mathbf{0} &  \mathbf{0} & \mathbf{T}_{Nx-1}^{\dagger} &  \mathbf{M}_{Nx}
\end{bmatrix}.
\end{equation}
Here, the $\mathbf{M}_i$ and $\mathbf{T}_i$ blocks are $2N_y \times 2N_y$ matrices which encode the couplings for both particle and hole degrees of freedom. $\mathbf{M}_i$ contains all couplings among sites belonging to the $i$th column, while $\mathbf{T}_i$ couples sites in the $i$th column with those in the $i+1$th column. One can thus view the system as quasi-one dimensional in the $x$-direction, with $\mathbf{T}_i$ serving as the ``hopping'' and $\mathbf{M}_i$ as the ``on-site energy.'' The diagonal blocks of $\mathbf{G}$ can then be obtained by a recursive algorithm with computational complexity $O(N_x N_y^3 )$ for each frequency at which the calculation is performed \cite{godfrin1991method,reuter2012efficient}. To be more specific, we introduce two sets of auxilliary matrices $\mathbf{A}_i$ and $\mathbf{B}_i$, all with dimension $2N_y \times 2N_y$ and defined recursively in the following way:
\begin{equation}
\mathbf{A}_i  = 
\begin{cases}
\mathbf{0} & \text{for } i = 1, \\
\mathbf{T}_{i-1}^{\dagger}\Bigl[(\omega + i\Gamma)\mathbf{1} - \mathbf{M}_{i-1} - \mathbf{A}_{i-1} \Bigr]\mathbf{T}_{i-1} & \text{for } 1 < i \leq N_x \\
\end{cases}
\end{equation}
and
\begin{equation}
	\mathbf{B}_i  = 
	\begin{cases}
		\mathbf{0} & \text{for } i = N_x, \\
		\mathbf{T}_{i}\Bigl[(\omega + i\Gamma)\mathbf{1} - \mathbf{M}_{i+1} - \mathbf{B}_{i+1} \Bigr]\mathbf{T}_{i}^{\dagger} & \text{for } 1 \leq i < N_x. \\
	\end{cases}
\end{equation}
The diagonal blocks of $\mathbf{G}$ are now given by the following expression:
\begin{equation}
\mathbf{G}_{ii}= \Bigl[(\omega + i\Gamma)\mathbf{1} - \mathbf{M}_{i} - \mathbf{A}_{i} - \mathbf{B}_{i} \Bigr].
\end{equation}
$\mathbf{G}_{ii}$ contains all Green's function elements such that $x_i = x_j$. From these diagonal blocks, one can extract the \emph{electron} LDOS from 
\begin{equation}
\rho(\mathbf{r}_i,\omega) = -\frac{1}{\pi} \operatorname{Im}G(\mathbf{r}_i,1 | \mathbf{r}_i,1 | \omega),
\label{eq:ldos}
\end{equation}
where only the ``particle-particle'' part of the Green's function is to be considered (hence $\sigma_i = 1$). This method allows us to make use of very large system sizes without making any approximations; in our calculations, we set $N_x = 1024$ and $N_y = 256$, resulting in minimal finite-size effects and extremely large fields of view that allow us to obtain accurate statistical correlations from different quantities that can be obtained from individual LDOS spectra. We fix the broadening parameter $\Gamma = 0.01$; we do not include any frequency-dependent renormalization of the self-energy.

\subsection{Order-Parameter Inhomogeneity}

The main objective of this paper is to examine the role of the length scale of order-parameter inhomogeneity on the LDOS and the various observables that can be extracted from it. It is important then to pick a model whose length scale is easily tuned, keeping everything else the same. There is no lack of models of order-parameter inhomogeneity which have appeared in the literature. One extensively studied model is the Yukawa-like model used by Nunner et al., in which off-plane impurities give rise to a modulated pairing interaction whose distance-dependence is that of a screened Coulomb potential \cite{nunner2005dopant,melikyan2006gap,nunner2006fourier,andersen2006thermodynamic}. While realistic in that one can use it to reproduce the gap distributions seen in experiment, the overall inhomogeneity length scale in this model is not easily tuned. Three length scales are present in the Yukawa problem: the distance between the impurity and the CuO$_2$ plane, the screening length, and the mean distance between impurities. The resulting overall length scale $l$ of the order parameter inhomogeneity then has a nontrivial dependence on the interplay among these three Yukawa length scales, making it not straightforward to adjust $l$.

Instead, we use as a basis the simpler model first studied by Fang et al. \cite{fang2006gap}. The model consists of a square patch of length $l$ within which the $d$-wave order parameter takes on a constant value, motivated by STS gap maps that show a patchy structure of the gap with what appears to be ``phase separation'' between small-gap and large-gap regions. (A similar square-patch model was used by Alvarez and Dagotto, but with random \emph{phases} instead of amplitudes on each patch \cite{alvarez2008fermi}.) Instead of considering a single isolated patch as Fang et al. did, we instead consider a ``many-patch'' version of this disorder model. The entire $N_x \times N_y$ system is subdivided into square regions, each with size $l \times l$. Let us first define $\Delta_{D,i}$ as the site-centered order parameter at site $i$ from which the bond order parameters (which enter the Hamiltonian) are derived. For each square patch, $\Delta_{D,i}$ is constant for all sites $i$ within that patch. The bond order parameter $\Delta_{ij}$ (which enters Eq.~\ref{eq:hamiltonian}) is then obtained from $\Delta_{D,i}$ via the following relation:
\begin{equation}
\Delta_{ij} = \frac{1}{4} \theta_{ij}\frac{\Delta_{D,i}+\Delta_{D,j}}{2}.
\label{eq:deltabond}
\end{equation}
Here, $\theta_{ij} = 1$ ($\theta_{ij} = -1$) if $i$ and $j$ are nearest-neighbor sites along the $x$- ($y$-) direction, and is zero otherwise. Note that with this choice of definition of $\Delta_{ij}$, the magnitude of the bond order parameter for bonds on domain walls between patches will be the average of $\Delta_D$ for the two patches. The additional factor of $\frac{1}{4}$ is needed to correctly normalize the bond order parameter.

The values of $\Delta_D$ on each patch are independent and identically distributed, with their values taken from a uniform distribution with mean $\overline{\Delta}$ and width $2\overline{\Delta}$. This choice of distribution ensures that a wide range of order parameter values is accessed. In particular, both large-gap patches (with $\Delta_D \approx 2\overline{\Delta}$) and small-gap patches (with $\Delta_D \approx 0$) coexist here. While our model for gap inhomogeneity does not appear to be similar to known gap distributions from STS experiments, it bears emphasizing that the gap extracted from experiment is not exactly the same as the underlying order parameter, and we make no attempt at fine-tuning the parameters to ensure that this model accurately reproduces the experimentally obtained distributions. Nevertheless, we show that in certain regimes, the resulting distribution of the gaps becomes similar to that seen in experiment \emph{without} any fine-tuning, the square shape of the underlying patches notwithstanding.

We also consider the maximally disordered case where the order parameter \emph{on every bond} is independent and identically distributed from a uniform distribution with mean $\frac{\overline{\Delta}}{4}$ and width $\frac{\overline{\Delta}}{2}$ (the $\frac{1}{4}$ prefactor again is needed for proper normalization of the bond order parameter). We call this the ``$l = 0$'' case as it is the extreme limit where no spatial correlations are present between the bond order parameters. Note that the smallest patch in our disorder model ($l=1$) still exhibits correlations among adjacent bond order parameters, as these are derived from $\Delta_D$ via Eq.~\ref{eq:deltabond}.

Because the objective of this paper is to examine the effect of the length scales of the patchy inhomogeneity, we neglect the addition of oscillatory components of the $d$-wave order parameter that have been identified in recent STS experiments on Bi-2212 \cite{hamidian2016detection,edkins2019magnetic,du2020imaging} and Bi-2223 \cite{zou2022particle}. This oscillatory component, which is found to be much smaller relative to the uniform component, has been identified as a ``pair-density wave'' in Bi-2212, but an alternative disorder-based mechanism involving quasiparticle scattering between antinodal states has been proposed for this which accounts for the particle-hole asymmetry of the oscillations in the coherence-peak location \cite{zou2022particle,gao2024pair}. We will not include these ingredients into the present paper, and only focus on the inhomogeneity of the \emph{background} (i.e., non-oscillatory) $d$-wave order parameter. 

The only parameters that we tune here are the size $l$ of the patches and the mean $\overline{\Delta}$; the latter sets the average coherence length $\xi$, and how large $l$ is compared with $\xi$ will turn out to be of significance to many observables. We show results primarily for $\overline{\Delta} = 0.16$ (with a coherence length $\xi \approx 4.53$ lattice spacings, this is a fairly accurate representation of mildly overdoped superconducting cuprates), but we also present calculations for $\overline{\Delta} = 0.32$ ($\xi \approx 2.27$) and $\overline{\Delta} = 0.08$ ($\xi \approx 9.06$).

\subsection{Microscopic Justification for the Order-Parameter Inhomogeneity Model}

Even though our model appears to be artificial as it involves square patches, it is intended to capture important aspects of the cuprates within the simplest model of inhomogeneity possible. We highlight two features of this model that (with the aid of hindsight) are central to the phenomenology of the cuprates. The first aspect is the presence of a length scale (in our model, the size of the patches $l$) that governs the range of spatial correlations of the order parameter inhomogeneity. The second aspect is an extremely broad distribution of order-parameter values.  In our treatment, both the inhomogeneity length scale and the width of the disorder are tunable parameters that we do not tie to a specific microscopic model. One can ask if these phenomenological assumptions are well-founded. In this section, we discuss a specific model that can give rise to the salient features of our chosen order-parameter inhomogeneity model, and discuss two possible microscopic realizations of this in the literature. 

We first note that one can obtain these order-parameter distributions by assuming the existence of a spatially varying attractive nearest-neighbor pairing interaction $V_{ij}$ (here, $i$ and $j$ label sites on the square lattice). In a self-consistent mean-field treatment, the bond order parameter $\Delta_{ij}$ is given by \cite{sulangi2025inhomogeneity}
\begin{equation}
	\Delta_{ij} = V_{ij}\langle c_{i\uparrow}c_{j\downarrow}\rangle.
	\label{eq:deltapairing}
\end{equation}
With a suitably chosen spatial profile of the pairing interaction $V_{ij}$, one can in principle reproduce any order-parameter inhomogeneity profile, including ours. We in fact show this in detail in a detailed follow-up work \cite{sulangifuture1}, where we demonstrate that broadly distributed square-patch inhomogeneity can arise within self-consistent mean-field theory, \emph{assuming} that the pairing interaction $V_{ij}$ itself is also broadly distributed and follows the square-patch inhomogeneity model. It is evident that a microscopic model underpinning our phenomenological one involves a \emph{spatially varying pairing interaction} treated self-consistently \cite{nunner2005dopant,melikyan2006gap,fang2006gap,andersen2006thermodynamic,romer2018raising,tromp2023puddle,sulangi2025inhomogeneity}. Further aspects of pairing-interaction inhomogeneity, including its effects on $T_c$ and the superfluid density, have previously been discussed in Refs. \cite{martin2005enhancement,zou2008effect,mishra2008sublattice} and will be expounded upon further in our upcoming work \cite{sulangifuture1}. 

It is helpful to recap why inhomogeneity in the \emph{pairing interaction} is likely to be present in the cuprates, and why this is our preferred starting point in our analysis. Nunner et al. find that many correlations seen among STS observables (e.g., the local spectral gap, the coherence-peak height, and dopant locations), as well as particle-hole symmetric local spectra, are reproduced nicely within mean-field theory if one assumes that the dopant atoms cause local modulations in the pairing interaction \cite{nunner2005dopant,melikyan2006gap}. Similarly, Fang et al. also find that the anticorrelation between the spectral gap size and the coherence peak height arises naturally from nanoscale inhomogeneity in the order parameter alone \cite{fang2006gap}. In addition, Sulangi et al. show that a good deal of the cuprate STS phenomenology is reproduced using the pairing-interaction inhomogeneity model of Nunner et al. treated within both self-consistent mean-field theory and time-dependent Ginzburg-Landau theory, with the latter demonstrating consistency with STS experiments done at high temperatures \cite{sulangi2025inhomogeneity}.

It is important to note that if one instead assumes a \emph{constant} pairing interaction and inhomogeneity in the chemical potential (with order-parameter inhomogeneity generated self-consistently), many of the aforementioned features are not easily reproduced within mean-field theory. In particular, Nunner et al. \cite{nunner2005dopant,melikyan2006gap} and Sulangi et al. \cite{sulangi2021correlations} find that within such a scenario, the LDOS near impurities are quite particle-hole asymmetric, in contrast with experiment. More importantly, for pure chemical-potential disorder to generate the order-parameter variations seen in experiment, this has to be strong enough that large local charge modulations \cite{nunner2005dopant}, subgap states, and gap filling \cite{pal2023simulating} are generated, which are not seen in STS. As gap filling is absent at low temperatures in the cuprates throughout most of the doping ranges we are concerned with, it is highly unlikely then that the inhomogeneity in the order parameter is the result of \emph{mean-field} backreaction of the order parameter on strong on-site potentials. No evidence for kinks and a low-energy--high-energy dichotomy akin to experiment is seen in self-consistent mean-field models involving only on-site potential disorder \cite{pal2023simulating}.

Inhomogeneity in the hopping amplitudes has been considered by Lee et al. \cite{lee2016cold}. Here, they assume a uniform pairing interaction, but the hoppings are modulated (with some correlation length) in such a way that a ``nematic glass'' is formed; their treatment is entirely mean-field in character. The generated superconducting order parameter is spatially extended, with the length scale set by that of the underlying nematic inhomogeneity. It is not clear from their analysis if the correlations among STS observables are reproduced within their model. However, they find a version of the low-energy--high-energy dichotomy in their calculated LDOS spectra. However, they do not appear to be able to reproduce kinks in the LDOS within their model. Instead, they find that the LDOS is homogeneous at low energies below the average gap energy, and inhomogeneous above it, a result \emph{not} similar to what is seen in STS experiments. It is not known if this discrepancy is because the resulting order-parameter variation is too narrow (compare with our results in Appendix A for some results on the order parameter distribution width-dependence of the LDOS), or if this is due to effects peculiar to the normal-state properties of their model of glassy nematicity.

What can cause an inhomogeneous pairing interaction to form? It is likely that this is due to the correlated nature of the cuprates. This is the point of view that R\o mer et al. \cite{romer2012local,romer2018raising} and Chakraborty et al. \cite{chakraborty2014fate,chakraborty2017effects,chakraborty2017pairing} adopt. R\o mer et al. \cite{romer2012local,romer2018raising} make use of a real-space formulation of spin fluctuations (treated within the random-phase approximation) to derive an effective pairing interaction, which owing to self-consistency will be sensitive to the presence of disorder in the on-site potential and in the hoppings (in their paper, they only consider site disorder in the form of nonmagnetic impurities). They then feed this inhomogeneous pairing interaction into a self-consistent Bogoliubov-de Gennes calculation, yielding an inhomogeneous $d$-wave superconductor. The resulting spectral gap maps with randomly distributed impurities are strikingly patchy and are reminiscent of experiment. This work represents a proof of principle that enhanced pairing can be ``nucleated'' in the vicinity of pointlike impurities, thereby providing a concrete microscopic model underlying the one used in Nunner et al.  In particular, the enhancement of the spectral gap was found to extend several lattice spacings around a single isolated impurity, and particularly strong enhancements in turn can be generated by rare diagonal ``dimer'' impurity configurations. These allow for the generation of broadly distributed and spatially extended order parameter inhomogeneity; the effects are even more pronounced the closer one gets to the magnetic instability point. A similar observation is made by a first-principles study which finds that the pairing interaction is \emph{enhanced} in the vicinity of a oxygen dopant in Bi-2212 \cite{foyevtsova2010modulation}.

Chakraborty et al. \cite{chakraborty2014fate,chakraborty2017effects,chakraborty2017pairing} on the other hand obtain a Gutzwiller-renormalized normal-state Hamiltonian. The Gutzwiller factors that encode the effects of strong correlations then enter into the self-consistent expression for the superconducting order parameter; a spatially dependent pairing interaction is generated as a result. It is worth noting that in the correlated calculation, they find the inhomogeneity in the resulting order parameter to be less sensitive to the underlying disorder than in the uncorrelated case. It would appear that within the Gutzwiller-renormalized theory, much stronger disorder needs to be present to reproduce the broad gap distributions seen in experiment \cite{chakraborty2017effects}. Correlation effects as modeled by the Gutzwiller-renormalized theory tend to keep the LDOS surprisingly uniform even in the presence of strong disorder, but this may partially be due to the small length scale (on the order of the Fermi wavelength $k_F^{-1}$) of the ensuing order-parameter inhomogeneity, which would give rise to a largely uniform quasiparticle response. (For comparison and corroborating evidence from a mean-field calculation, please see our $l < \xi$ results in the next section.)

In any case, the differences between Ref. \cite{romer2018raising} and Refs. \cite{chakraborty2014fate,chakraborty2017effects,chakraborty2017pairing} highlight the sensitivity of the resulting observables to the particulars of the correlation model used. This is especially so when the ultimate source of inhomogeneity is not settled with any sort of definiteness. We have therefore opted to treat inhomogeneity entirely within a phenomenological model, and we have not included correlation effects in the normal-state Hamiltonian, similar to Nunner et al. \cite{nunner2005dopant,melikyan2006gap}, Fang et al. \cite{fang2006gap}, and Tromp et al. \cite{tromp2023puddle}. This ``back-to-basics'' correlation-agnostic approach ultimately allows us to isolate what \emph{non-correlation-related} ingredients are necessary for the phenomenology of STS experiments on superconducting cuprates across a wide doping range, over which the strength of the electron-electron correlations presumably varies considerably.

\subsection{Order Parameter Proxies}

A central motivation of this paper is figuring out the extent to which the various observables that can be extracted from $dI/dV$ measurements can serve as accurate proxies for the true underlying superconducting order parameter. In much of the STS literature, the energy at which the superconducting coherence peak is found---the ``spectral gap''---has long served as a convenient proxy for the superconducting gap (on the other hand, it should be noted that many other STS papers interpret the spectral gap as the pseudogap energy). For a bulk superconductor in the clean limit, this correspondence is exact. However, in the presence of disorder, this is no longer automatically true. When chemical-potential disorder is present, the correlation between the spectral gap and the underlying order parameter is not strong, and in certain circumstances becomes very weak \cite{sulangi2021correlations}. As this paper focuses purely on order-parameter inhomogeneity, we reexamine the correlation between the spectral gap and the order parameter in detail, and in particular focus on the role played by the inhomogeneity length scale $l$ on this correlation.

We focus on two proxies for the local superconducting order parameter. The first one is simply the aforementioned spectral gap $\Delta_S$ and is obtained by finding the location of the positive-energy peak in the LDOS. (The negative-energy spectra are disregarded as these are contaminated by the presence of the Van Hove singularity right below the Fermi energy.) The second one is christened the ``low-energy gap'' $\Delta_L$ and is obtained by taking the slope of the positive-energy LDOS near zero energy. In a bulk clean $d$-wave superconductor, the slope near $\omega = 0$ is inversely proportional to the bulk order parameter, so one can use the former to obtain the latter. We use this relationship $d\rho/d\omega|_{\omega\approx 0^{+}} \propto 1/\Delta_L$ to obtain an estimate for $\Delta_L$ from the low-energy slope of the LDOS at each lattice site. To be more precise, because we use a finite frequency grid in evaluating the LDOS, we use the finite difference quotient to approximate the low-energy slope of the LDOS:
\begin{equation}
\frac{d\rho(\mathbf{r}_i,\omega)}{d\omega}|_{\omega\approx 0^{+}} \approx \frac{\rho(\mathbf{r}_i,\omega = 0.02) - \rho(\mathbf{r}_i,\omega = 0.01)}{0.01}.
\label{eq:ldosslope_estimate}
\end{equation}
Using a momentum-space calculation and the above estimate for the low-energy slope, we find that the uniform $d$-wave order parameter $\Delta_T$ and the low-energy slope (with $\Gamma = 0.01$ and the same band-structure parameters as used before) are inversely related via the following formula:
\begin{equation}
\Delta_L = \Delta_T \approx 0.008574 + 0.1243\times\frac{1}{\frac{d\rho(\mathbf{r}_i,\omega)}{d\omega}|_{\omega\approx 0^{+}}}.
\label{eq:leg_estimate}
\end{equation}
While this estimate is exact only for a homogeneous $d$-wave superconductor, we will nevertheless use Eqs.~\ref{eq:ldosslope_estimate} and~\ref{eq:leg_estimate} to estimate the low-energy gap $\Delta_L$ for each LDOS spectrum for the inhomogeneous case.

These two proxies are to be compared with the ``true'' superconducting order parameter $\Delta_T$, which we define in the following way:
\begin{equation}
\Delta_{T,i}= \sum_{\epsilon} \theta_{i,i+\epsilon} \Delta_{i,i+\epsilon}.
\label{eq:deltat}
\end{equation}
Here, $\theta_{ij}$ is defined similarly as before, and $\epsilon$ is an integer such that $i+\epsilon$ is one of the four nearest neighbors of site $i$. $\Delta_T$ is a measure of the strength of the local $d$-wave pairing at a particular site $i$ and is simply the sum of the four bond order parameters connecting site $i$ with its nearest neighbors, with the sign difference between $x$- and $y$-aligned bond order parameters taken into account. Note that $\Delta_T$ here is not the same as $\Delta_D$ defined earlier. These two will coincide inside the patches, but will differ in value at the edges. 

We also retain information about the coherence-peak height $\rho_C(\mathbf{r}_i) = \rho(\mathbf{r}_i, \omega = +\Delta_{S,i})$ as its correlations with the order-parameter proxies can potentially provide valuable clues about the nature of the order-parameter inhomogeneity in real-world cuprates. In particular, it is known that $\rho_C$ and $\Delta_S$ are \emph{negatively} correlated for many cuprates studied by STS such as Bi-2212 and Bi-2201 \cite{lang2002imaging,mcelroy2005coincidence,fang2006gap,alldredge2008evolution,zou2022particle}, so this provides an important experimental constraint on various theoretical models of order-parameter inhomogeneity.

\section{Small Patches ($l < \xi$)}

\begin{figure}[t]
	\centering
	\includegraphics[width=0.5\textwidth]{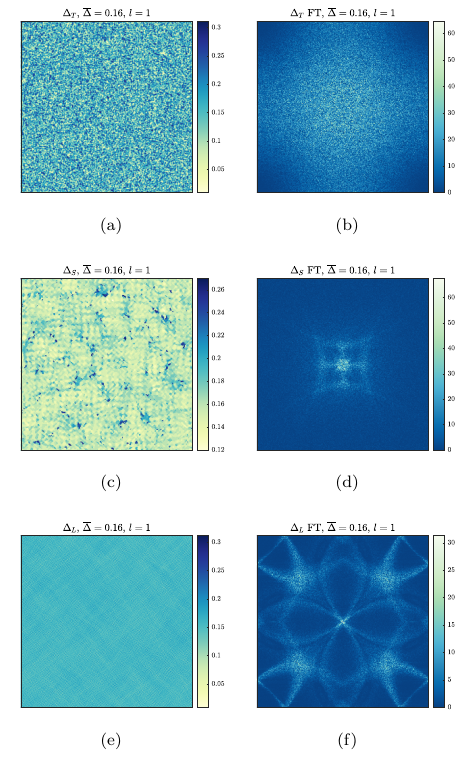}
	\caption{Plots of $\Delta_T$ (a), $\Delta_S$ (c), and $\Delta_L$ (e), and the absolute value of their respective Fourier transforms (b, d, and f) for patch size $l = 1$. Only one disorder realization is presented. Shown here are quantities taken from the middlemost $256 \times 256$ segment of the full $1024 \times 256$ system.}
	\label{fig:centeredplots_1}
\end{figure}

\begin{figure}[t]
	\centering
	\includegraphics[width=0.5\textwidth]{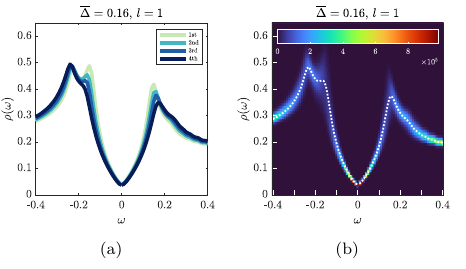}
	
	\caption{Left: Plots of the binned average local density of states for $l = 1$. Each LDOS spectrum is binned according to its spectral gap (with four bins in all), and all spectra in a given bin are then averaged over. Right: Plot of the distribution of the local density of states as a function of energy (heat map) and the average local density of states (dashed white line) for $l = 1$. The spectra are binned according to the energy, and histograms for each energy are taken. The counts per energy bin are shown as a heat map. For both these plots, 4 disorder realizations and a total of 1,048,576 individual spectra are used in this plot.}
	\label{fig:ldosbygap_ldoshist_1}
\end{figure}

\begin{figure}[t]
	\centering
	\includegraphics[width=0.5\textwidth]{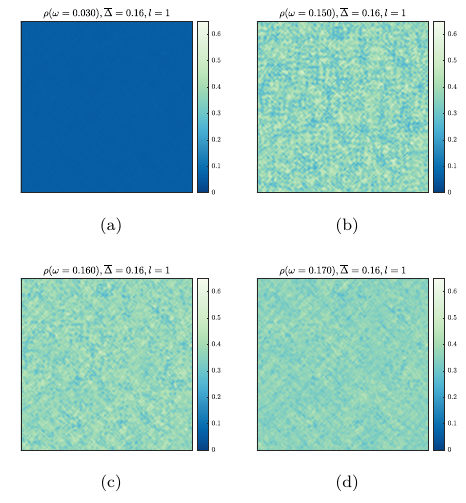}
	\caption{Plots of the LDOS at different frequencies for patch size $l = 1$. The same disorder realization and field of view as in Fig.~\ref{fig:centeredplots_1} are used here.}
	\label{fig:ldosplots_1}
\end{figure}

\begin{figure}[t]
	\centering
	\includegraphics[width=0.44\textwidth]{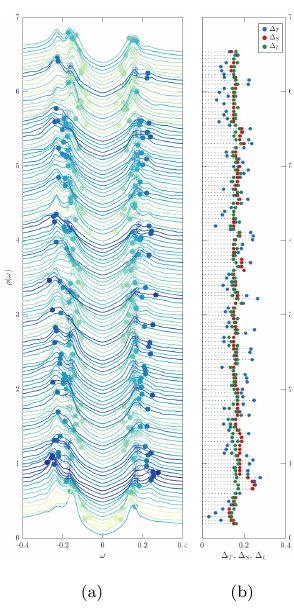}
	
	\caption{Left: Plots of the local density of states along a straight line through the middle of the sample from $(516,64)$ to $(516,192)$ for $l = 1$. The filled circles indicate the position of the underlying order parameter $\Delta_T$, while open circles and asterisks indicate the spectral gap $\Delta_S$ and low-energy gap $\Delta_L$, respectively. Only a single disorder realization is shown here. The plots are colored according to the local value of $\Delta_T$. Right: Plots of $\Delta_T$ (blue), $\Delta_S$ (red), and $\Delta_L$ (green) along the same linecut.}
	\label{fig:linecuts_1}
\end{figure}

\begin{figure}[t]
	\centering
	\includegraphics[width=0.5\textwidth]{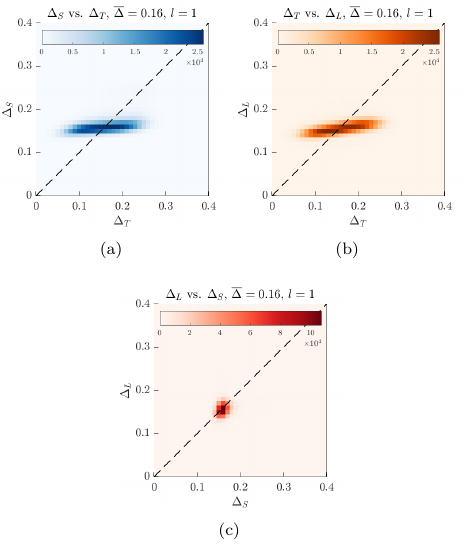}

	\caption{Plots of the two-dimensional histograms for a) $\Delta_T$ and $\Delta_S$; b) $\Delta_T$ and $\Delta_L$; and c) $\Delta_S$ and $\Delta_L$ for $l = 1$. Included here are data points from four disorder realizations corresponding to 1,048,576 individual LDOS spectra.}
	\label{fig:histogram_opsgleg_1}
\end{figure}

\begin{figure}[t]
	\centering
	
	\includegraphics[width=0.5\textwidth]{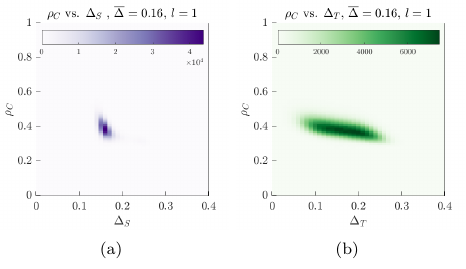}
	
	\caption{Plots of the two-dimensional histograms for a) $\Delta_S$ and $\rho_C$ and b) $\Delta_T$ and $\rho_C$ for $l = 1$. Included here are data points from four disorder realizations corresponding to 1,048,576 individual LDOS spectra.}
	\label{fig:histogram_opsgcph_1}
\end{figure}

We first turn our attention to small patches characterized by $l < \xi$. To exemplify this, we show in Fig.~\ref{fig:centeredplots_1} real-space and Fourier-transformed plots of $\Delta_T$, $\Delta_S$, and $\Delta_L$ for a $256 \times 256$ subset of the system for the case where the mean order parameter is $\overline{\Delta} = 0.16$ and the patch size is $l = 1$. As mentioned earlier, for $\overline{\Delta} = 0.16$, the coherence length is $\xi \approx 4.53$, so the patches are parametrically smaller than the coherence length. It can be seen from the real-space plot of the order parameter $\Delta_T$ (Fig.~\ref{fig:centeredplots_1}a) that it has a broad distribution ranging from $\Delta_T \approx 0$ to $\Delta_T \approx 0.3$, and that its modulations are short-ranged, with $\Delta_T$ on nearby sites largely uncorrelated with one another, except for rare regions where multiple patches with similar values are close to one another. The distribution of $\Delta_T$ is broad in that both regions with very small gaps and regions with very large gaps are present, often in close proximity with one another. In contrast, the spectral gap $\Delta_S$ (Fig.~\ref{fig:centeredplots_1}c) is comparatively narrowly distributed around $0.16$, except for a very small number of regions whose value clusters around $0.25$. The latter effect is due to the Van Hove singularity (VHS); in the \emph{clean} limit with $\overline{\Delta} = 0.16$, the VHS gives rise to a small shoulder at $\omega \approx 0.25$ above the positive-energy coherence peak at $\omega = 0.16$. In the inhomogeneous setting, it appears that interference effects can cause this shoulder to mix with the coherence peak, leading to the formation of a peak where the shoulder should be. These clusters tend to form at large-$\Delta_T$ regions, but this effect appears to be rare as only a small number of large-$\Delta_T$ regions exhibit this effect at all.

The real-space plot of $\Delta_S$ is characterized by weak modulations with a wavelength of around a few lattice sites. If one closely examines the $\Delta_S$ and $\Delta_T$ plots side by side, it can be seen that the two maps do not look similar to one another at all. While some large-$\Delta_T$ areas coincide with areas where $\Delta_S$ is large, the periodic modulations seen in $\Delta_S$ do not at all correspond to any modulations seen in $\Delta_T$. These periodic modulations are instead due to QPI arising from the presence of short-range disorder in the pairing (or ``off-diagonal'') channel. The Fourier transform of the $\Delta_S$ map (Fig.~\ref{fig:centeredplots_1}d) confirms this: the dominant wavevectors here are precisely the scattering wavevectors responsible for QPI near the $d$-wave gap edge. This explanation is not altogether surprising. In the $l < \xi$ regime, the order-parameter inhomogeneity can be regarded as being in the pointlike limit in which the allowed scattering processes are unrestricted except for the requirement that they connect quasiparticle states for which the order parameter has the \emph{same} sign. In the $l = 1$ case, the modulations in the LDOS at $\omega \approx 0.16$ themselves form the modulations in the $\Delta_S$ maps (recall that $\Delta_S$ is the location of the positive-energy coherence peak). A close examination of Fig.~\ref{fig:centeredplots_1}d shows that the dominant scattering wavevectors giving rise to the modulations in the $\Delta_S$ map are the so-called sign-preserving ``$\mathbf{q}$-vectors'' well-known from the octet-model description of QPI: $\mathbf{q}_1$, $\mathbf{q}_4$, and $\mathbf{q}_5$ \cite{nunner2006fourier,pereg2008magnetic,sulangi2017revisiting}.

$\Delta_L$ (Fig.~\ref{fig:centeredplots_1}e) meanwhile can be seen to be very narrowly distributed around 0.16 as well, and does not reflect the comparatively large spatial variations in the underlying order parameter $\Delta_T$. Modulations can be seen here as well, but these are evidently very weak, and have a different characteristic length scale that those seen in $\Delta_S$. The Fourier transform of $\Delta_L$ (Fig.~\ref{fig:centeredplots_1}f) shows that these modulations arise from QPI as well, but this time at low energies near $\omega \approx 0$. The Fourier transforms (with visible QPI scattering wavevectors) of the two order-parameter proxies in Figs.~\ref{fig:centeredplots_1}d and~\ref{fig:centeredplots_1}f are in striking contrast to the Fourier transform of the true order parameter $\Delta_T$ (Fig.~\ref{fig:centeredplots_1}b), which is by and large featureless.

The homogeneous nature of the superconducting state with $l = 1$ throughout the entire energy range $E \in [-\overline{\Delta}, \overline{\Delta}]$ is further illustrated in Fig.~\ref{fig:ldosbygap_ldoshist_1}. In Fig.~\ref{fig:ldosbygap_ldoshist_1}a, we collect 1,048,576 individual LDOS spectra and sort these by their spectral gap $\Delta_S$; these spectra are then divided into four bins, and all spectra within a single bin are then averaged over. (Only four bins are used here because of the extreme narrowness of the $\Delta_S$ distribution.) It can be seen that the differences between the spectra from the lowest bin (with the smallest $\Delta_S$) and those from the largest bin (with the largest $\Delta_S$) are small, with only a minimal shift in the coherence-peak positions. The main difference is in the height of the peaks, with the small-$\Delta_S$ lineshapes showing taller and sharper coherence peaks at both positive and negative energies, and the large-$\Delta_S$ ones displaying shorter and more rounded peaks. In the $l < \xi$ regime exemplified here, the bin-averaged $\Delta_S$ and $\rho_C$ are negatively correlated with one another, as in experiment. However, the LDOS spectra are very similar to one another and do not display the diverse variety of lineshapes that are seen in STS. 

Fig.~\ref{fig:ldosbygap_ldoshist_1}b shows a two-dimensional histogram of the LDOS \emph{values} across the same ensemble of spectra shown in the previous plot, with the disorder- and system-averaged LDOS shown in white. For each value of the frequency $\omega$, the distribution of the LDOS is obtained, and the density of each possible LDOS value is shown in Fig.~\ref{fig:ldosbygap_ldoshist_1}b. It can be seen that the LDOS distribution does not deviate much from the average LDOS. This is especially true at low energies. As the energy is increased, one can see that the distribution becomes broader as $ \to \pm \overline{\Delta}$, but the values remain within a narrow region surrounding the average LDOS. 

More insight into the homogeneous nature of the states comes by way of plotting the LDOS itself (Fig.~\ref{fig:ldosplots_1}) taken from the same field of view as in Fig.~\ref{fig:centeredplots_1} at four representative energies, which we take to be: $\omega = 0.03$ (Fig.~\ref{fig:ldosplots_1}a), to represent the low-energy LDOS; $\omega = 0.15$ and $\omega = 0.17$ (Figs.~\ref{fig:ldosplots_1}b and d), which are the spectral gaps of the LDOS averaged over the \emph{first} and \emph{last} bins, respectively, in Fig.~\ref{fig:ldosbygap_ldoshist_1}b; and $\omega = 0.16$, which corresponds to the average superconducting order parameter in the system. The latter three energies are chosen to exemplify the high-energy states where inhomogeneity is strongest. The low-energy LDOS (Fig.~\ref{fig:ldosplots_1}a) is very featureless and does not contain any strong variations reflective of the underlying order parameter. As for the high-energy plots (Figs.~\ref{fig:ldosplots_1}b-d), these all are very similar to one another in that they feature modulations due to QPI. These modulations are particularly pronounced for $\omega = 0.15$ (Fig.~\ref{fig:ldosplots_1}b), but they appear to weaken as energy is increased from this point. These LDOS patterns are consistent with QPI due to weak, narrowly distributed order-parameter inhomogeneity as previously studied by Sulangi et al. \cite{sulangi2017revisiting}, but it is nevertheless striking that the same patterns are still present here even when the width of the order-parameter disorder distribution is parametrically much larger than in the case previously studied. This may be attributed to the fact that even with the relatively large width of the order-parameter inhomogeneity distribution we have used (on the order of the mean superconducting order parameter itself), it is still weak on the overall scale set by the bandwidth ($\approx t$), and therefore what results is still QPI with very clear and visible modulations.

In Fig.~\ref{fig:linecuts_1}a, we show a plot of the LDOS across a single cut through a sample alongside plots of $\Delta_T$. The individual LDOS lineshapes do not vary much moving across the cut, especially at low energies. At high energies, some inhomogeneity in the LDOS clearly appears, but the peak positions do not vary much. One can see that the position of the coherence peak often does not match $\Delta_T$ for that particular lattice site; the latter varies very quickly as a function of position, whereas the LDOS response is much more uniform. This is also shown in Fig.~\ref{fig:linecuts_1}b, which shows both $\Delta_T$, $\Delta_S$ and $\Delta_L$ along the same linecut. $\Delta_T$ oscillates dramatically from site to site, while $\Delta_S$ and $\Delta_L$ do not deviate considerably from $\overline{\Delta} \approx 0.16$. Peaks and dips can be seen in the profiles of $\Delta_S$ and $\Delta_L$, but these do not appear to correspond clearly to peaks and dips in $\Delta_T$ except for only a few instances where clusters of small- or large-$\Delta_T$ regions give rise to a dip or peak, respectively, in $\Delta_S$ and $\Delta_L$. 

The correlations among the various order-parameter measures can be seen in Fig.~\ref{fig:histogram_opsgleg_1}, which shows two-dimensional histograms for these quantities taken from all spectra we generated and puts on solid footing the observations we had made earlier from the real-space plots. In Fig.~\ref{fig:histogram_opsgleg_1}a, we show the histogram for $\Delta_T$ and $\Delta_S$. It can be seen that while the distribution of $\Delta_T$ is broad, the resulting spectral gaps are much more narrowly distributed, with barely any variation even as the underlying order parameter fluctuates greatly. The same results can be seen for the two-dimensional histograms of $\Delta_T$ and $\Delta_L$ (Fig.~\ref{fig:histogram_opsgleg_1}b): the distribution of $\Delta_L$ is much narrower than that of $\Delta_T$. Curiously, when we look at the 2D histogram of $\Delta_S$ and $\Delta_L$ (Fig.~\ref{fig:histogram_opsgleg_1}c), we find that it consists of a highly localized dot around $\Delta_S \approx \Delta_L \approx 0.16$. Evidently, both low- and high-energy measures of the order parameter are identically highly homogeneous and do not reflect the strongly fluctuating nature of $\Delta_T$. In Fig.~\ref{fig:histogram_opsgcph_1}, we plot the correlations of $\Delta_T$ and $\Delta_S$ with the coherence-peak height $\rho_C$; here it can be seen that both $\Delta_S$ and $\Delta_T$ are negatively correlated with $\rho_C$.

It is remarkable that the spectra remains homogeneous at both low and high energies even in the presence of a highly inhomogeneous order-parameter background. The reason for this in the $l < \xi$ case is that the quasiparticle response will be washed out over a coherence volume $\xi^2$, so any variations of the order parameter that are smaller than $\xi$ will be invisible in any observables that are directly derived from quasiparticle properties such as the LDOS. This can be seen again in Figs.~\ref{fig:centeredplots_1} and~\ref{fig:linecuts_1}: the extremely small-$\Delta_T$ regions do not exhibit a gapless spectrum, and instead show a finite $\Delta_S \approx \overline{\Delta}$. The superconducting quasiparticles in effect see a mostly homogeneous background with order parameter $\overline{\Delta}$ but are nevertheless scattered elastically by order-parameter inhomogeneity, which itself is \emph{weak} on the scale set by the nearest-neighbor hopping amplitude $t$. The effect of this scattering will be to broaden the spectral functions at higher energies. This is corroborated by Fig.~\ref{fig:ldosbygap_ldoshist_1}b: the LDOS at $\omega \approx 0.16$ is more broadly distributed than at lower energies, and consequently its average LDOS is lower than it would be in the absence of inhomogeneity. This is similar to what weak-impurity scattering in the Born limit predicts \cite{lee2018optical}, and suggests that the physics at $l < \xi$ is determined much more strongly by Born scattering (albeit in the ``off-diagonal'' channel) than in the other cases we will consider shortly, where the scattering is of a decidedly different nature. 

\section{Intermediate-Sized Patches ($l \approx \xi$)}

\begin{figure}[t]
	\centering
	\includegraphics[width=0.5\textwidth]{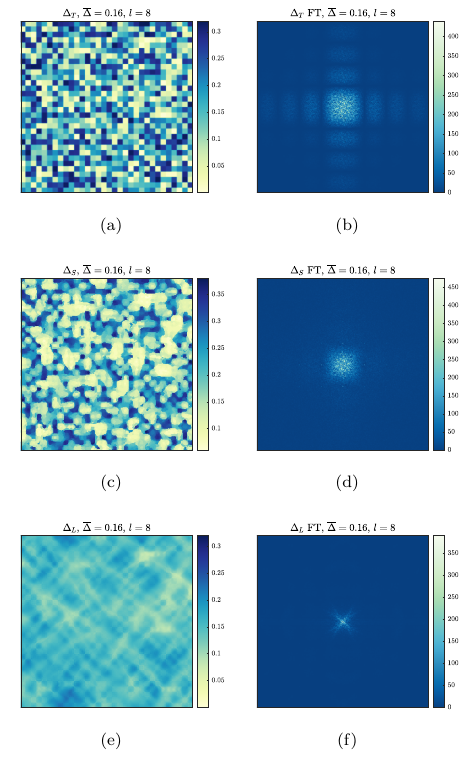}
	\caption{Plots of $\Delta_T$ (a), $\Delta_S$ (c), and $\Delta_L$ (e), and the absolute value of their respective Fourier transforms (b, d, and f) for patch size $l = 8$. Only one disorder realization is presented. Shown here are quantities taken from the middlemost $256 \times 256$ segment of the full $1024 \times 256$ system.}
	\label{fig:centeredplots_8}
\end{figure}

\begin{figure}[t]
	\centering
	\includegraphics[width=0.5\textwidth]{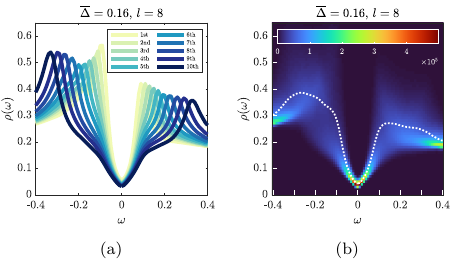}
	
	\caption{Left: Plots of the binned average local density of states for $l = 8$. Each LDOS spectrum is binned according to its spectral gap (with 10 bins in all), and all spectra in a given bin are then averaged over. Right: Plot of the distribution of the local density of states as a function of energy (heat map) and the average local density of states (dashed white line) for $l = 8$. The spectra are binned according to the energy, and histograms for each energy are taken. The counts per energy bin are shown as a heat map. For these two plots, 4 disorder realizations and a total of 1,048,576 individual spectra are used in this plot.}
	\label{fig:ldosbygap_ldoshist_8}
\end{figure}

\begin{figure}[t]
	\centering
	\includegraphics[width=0.5\textwidth]{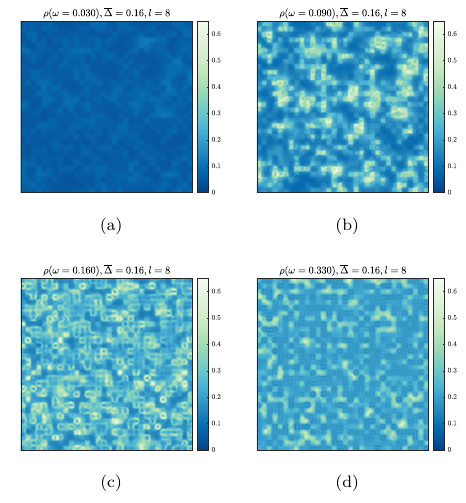}
	\caption{Plots of the LDOS at different frequencies for patch size $l = 8$. The same disorder realization and field of view as in Fig.~\ref{fig:centeredplots_8} are used here.}
	\label{fig:ldosplots_8}
\end{figure}

\begin{figure}[t]
	\centering
	\includegraphics[width=0.44\textwidth]{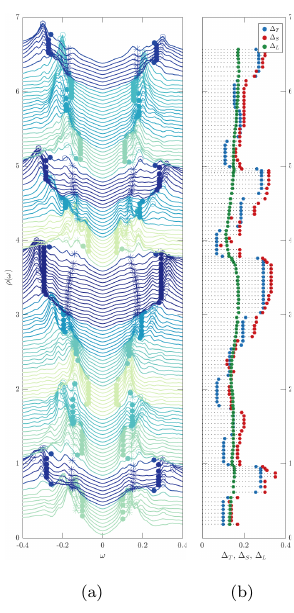}
	
	\caption{Left: Plots of the local density of states along a straight line through the middle of the sample from $(516,64)$ to $(516,192)$ for $l = 8$. The filled circles indicate the position of the underlying order parameter $\Delta_T$, while open circles and asterisks indicate the spectral gap $\Delta_S$ and low-energy gap $\Delta_L$, respectively. Only a single disorder realization is shown here. The plots are colored according to the local value of $\Delta_T$. Right: Plots of $\Delta_T$ (blue), $\Delta_S$ (red), and $\Delta_L$ (green) along the same linecut.}
	\label{fig:linecuts_8}
\end{figure}

\begin{figure}[t]
	\centering
	\includegraphics[width=0.5\textwidth]{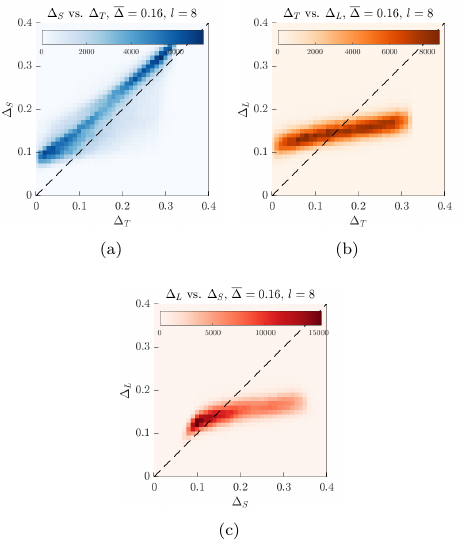}
	
	\caption{Plots of the two-dimensional histograms for a) $\Delta_T$ and $\Delta_S$; b) $\Delta_T$ and $\Delta_L$; and c) $\Delta_S$ and $\Delta_L$ for $l = 8$. Included here are data points from four disorder realizations corresponding to 1,048,576 individual LDOS spectra.}
	\label{fig:histogram_opsgleg_8}
\end{figure}

\begin{figure}[t]
	\centering
	\includegraphics[width=0.50\textwidth]{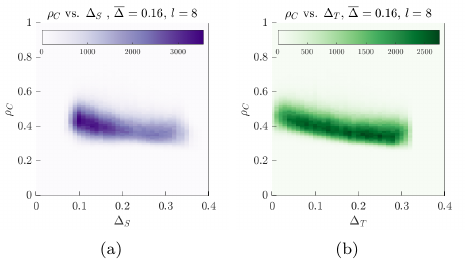}
	
	\caption{Plots of the two-dimensional histograms for a) $\Delta_S$ and $\rho_C$ and b) $\Delta_T$ and $\rho_C$ for $l = 8$. Included here are data points from four disorder realizations corresponding to 1,048,576 individual LDOS spectra.}
	\label{fig:histogram_opsgcph_8}
\end{figure}

We next consider intermediate-sized patches with $l \approx \xi$. Ths is the regime of most physical interest, as the cuprate superconductors are known to have both short coherence lengths on the order of a few lattice spacings and patchy gap inhomogeneity characterized by similar length scales. We take as our example the case of square patches with $l = 8$, and we again set $\overline{\Delta} = 0.16$ (with $\xi \approx 4.53$). Real-space and Fourier-transformed plots of $\Delta_T$, $\Delta_S$, and $\Delta_L$ are shown in Fig.~\ref{fig:centeredplots_8}; as with the $l = 1$ example, a $256 \times 256$ field of view is used in these plots.

The real-space $\Delta_T$ map (Fig.~\ref{fig:centeredplots_8}a) shows that $\Delta_T$ is broadly distributed; regions with very small $\Delta_T$ coexist with regions where $\Delta_T$ is very large. The effect of this extended form of inhomogeneity in the order parameter in the spectral gap $\Delta_S$ (Fig.~\ref{fig:centeredplots_8}c) is striking: $\Delta_S$ has a broad distribution, like $\Delta_T$, but looks far more splotchy than $\Delta_T$; the sharp square-shaped features seen in $\Delta_T$ are not present in these plots. Instead, $\Delta_S$ looks like a coarse-grained version of $\Delta_T$. It can be seen by eyeballing these two plots side by side that clusters of square patches where $\Delta_T$ is small show up in the $\Delta_S$ map as a spatially extended cluster where the spectra gap is small. An analogous observation can be made for the large-$\Delta_T$ clusters, but it can be seen that these show up in $\Delta_S$ as large-gap clusters that look smaller in spatial extent compared to the small-gap regions. While a clear correlation between $\Delta_T$ and $\Delta_S$ can be seen in these real-space maps, the correspondences are never exact between the two quantities. Importantly, it should be noted that while there exist regions where $\Delta_T \approx 0$, the spectral gap $\Delta_S$ never vanishes throughout the entire sample---the whole system exhibits a nonzero spectral gap, with the minimum spectral gap around $\approx 0.08$.

$\Delta_S$ no longer exhibits the periodic modulations seen in the $l  = 1$ case. This serves to highlight a key difference between this case and the $l < \xi$ case considered earlier: the nature of the scattering of the Bogoliubov quasiparticles off of the order-parameter inhomogeneity is different for both of these cases. Because $l$ is now a few lattice sites long, only small-momentum scattering occurs. This means that order-parameter-induced QPI is a weaker effect for $l \approx \xi$. This can be seen explicitly by examining the Fourier transforms of $\Delta_T$ and $\Delta_S$ (Figs.~\ref{fig:centeredplots_8}b and d). The Fourier transform of $\Delta_T$ shows that strong intensities are localized around a relatively small portion of the Brillouin zone centered around $\mathbf{k}=0$, in contrast with what is seen for $l = 1$ where the Fourier transform is featureless across the entire Brillouin zone; the resulting scattering off of the order-parameter inhomogeneity is primarily small-momentum and small-angle in nature. Higher harmonics are present in the Fourier transform of $\Delta_T$ owing to the uniformly square shape and size of the patches. The Fourier transform of $\Delta_S$ also shows similar characteristics: the intensity is largely localized near zero momentum, and the sharp, QPI-derived features seen in Fig.~\ref{fig:centeredplots_1}d for $l  = 1$ are no longer seen. However, the higher harmonics seen in the Fourier transform of $\Delta_T$ are no longer visible here as the patches in the $\Delta_S$ maps are more irregularly shaped.

In contrast to the inhomogeneous nature of the $\Delta_S$ plots, the low-energy gap $\Delta_L$ (Fig.~\ref{fig:centeredplots_1}e) remains very much homogeneous, hardly deviating from $\Delta_L \approx 0.16$. Only faint traces of the large variation in $\Delta_T$ can be found in $\Delta_L$. Regions where many large-$\Delta_T$ patches are in close proximity to one another will give rise to a larger $\Delta_L$, and vice versa, but these effects are extremely faint and certainly much more muted than in the case of $\Delta_S$. The Fourier transform of $\Delta_L$ (Fig.~\ref{fig:centeredplots_1}f) features strong signals only at small wavevectors, in contrast to the $l = 1$ case, indicating that for $l \approx \xi$, QPI even at low energies is affected by the presence of spatially extended order-parameter inhomogeneity, with the only real-space modulations present arising from small-momentum scattering processes. As $\Delta_L$ is a low-energy proxy for the superconducting order parameter and $\Delta_S$ a high-energy one, the difference in their behavior for $l = 8$ stands in contrast to that seen in the $l  = 1$ case, for which both order-parameter proxies are highly homogeneous (with the only differences stemming from the differing nature of QPI at low and high energies). The dichotomy between $\Delta_L$ and $\Delta_S$ suggests a fundamental difference between the low-energy and high-energy states of an inhomogeneous $d$-wave superconductor when the inhomogeneity length scale $l \approx \xi$.

To shed light on this apparent dichotomy, we plot in Fig.~\ref{fig:ldosbygap_ldoshist_8}a the bin-averaged LDOS (with the sorting of spectra done according to the size of the spectral gap, as was done for the $l  =1$ case, but this time with 10 bins, owing to the wide distribution of $\Delta_S$). Perhaps the most striking feature of this plot is the fact that all LDOS spectra taken from the ten bins look identical at low energies. These spectra start to deviate from one another at a characteristic energy $\Delta_K \approx 0.09$, which roughly corresponds to the spectral gap $\Delta_S$ of the first bin (which contains the spectra with the smallest spectral gaps). The spectra with the largest gaps feature particularly pronounced kinks which mark the transition from a homogeneous state at low energies to a highly inhomogeneous one at higher energies. On the other hand, the small-gap spectra show no kinks. In this $l \approx \xi$ regime, the heights of the coherence peaks of the bin-averaged spectra are inversely correlated with the corresponding $\Delta_S$ of these spectra. 

The contrast between the low- and high-energy states is further shown in  Fig.~\ref{fig:ldosbygap_ldoshist_8}b, which shows the two-dimensional histogram of the LDOS values alongside the system- and disorder-averaged DOS, similar to what we had plotted for the $l = 1$ case. At low energies, the LDOS distribution is sharp and well-defined, but as energy is increased beyond $\Delta_K \approx 0.09$, it spreads out and becomes broadly distributed (at fixed $\omega$). The LDOS distribution becomes sharp again only at the highest energies $\omega \approx 0.4$ beyond the highest $\Delta_T$. The broad distribution of the values of the LDOS at energies $\Delta_K < \omega < \text{max}(\Delta_T)$ has consequences for the \emph{average} LDOS (averaged over the entire system and over disorder realizations): a sharp coherence peak is no longer present, and in its place is a much more rounded peak, reflecting the presence of a broad range of $\Delta_S$ values at which the various LDOS spectra peak.

The real-space plots of the LDOS itself (Fig.~\ref{fig:ldosplots_8}) show how the underlying order-parameter inhomogeneity manifests itself in the quasiparticle states as frequency is increased. As in the $l = 1$ case, here we plot the LDOS taken from the same field of view as Fig.~\ref{fig:centeredplots_8} and at four representative frequencies: $\omega = 0.03$ (Fig.~\ref{fig:ldosplots_8}a, to exemplify the low-energy regime); $\omega = 0.09$ and $\omega = 0.33$ (Figs.~\ref{fig:ldosplots_8}b and d, corresponding to the spectral gaps of the first and last bins in Fig.~\ref{fig:ldosbygap_ldoshist_8}a); and $\omega = 0.16$ (Fig.~\ref{fig:ldosplots_8}c, corresponding to the average value of the superconducting order parameter). As with the $l = 1$ case, at low energies (Fig.~\ref{fig:ldosplots_8}a), the LDOS is largely featureless, with only small variations present. While it can be seen from comparing this plot with the $\Delta_T$ map (Fig.~\ref{fig:centeredplots_8}a) that regions with relatively large LDOS correspond to small-$\Delta_T$ regions, the overall scale of the variation remains very small. The story is different when we now look at the system at $\omega = 0.09$ (Fig.~\ref{fig:ldosplots_8}b), which is the energy at which the first bin of LDOS spectra exhibits a sharp spectral peak. We see here that the LDOS has become very inhomogeneous, with a substantial fraction of the system showing a very large density of states. However, large portions of the system still exhibit a small LDOS. A good correspondence can be seen between the large-LDOS regions of Fig.~\ref{fig:ldosplots_8}b and the small-$\Delta_T$ regions of Fig.~\ref{fig:centeredplots_8}a, and vice versa. It is worth remarking that both large-LDOS and small-LDOS regions exhibit no discernible modulations at this frequency, and that the square shape of the underlying small-$\Delta_T$ patches is sharply reproduced by the large-LDOS regions. 

At $\omega = 0.16$ (Fig.~\ref{fig:ldosplots_8}c), we can see that the large-LDOS regions are no longer easily identified with any particular set of patches in Fig.~\ref{fig:ldosplots_8}b, and no longer retain the characteristic square shape of the underlying order-parameter patches, appearing more irregularly shaped instead. In addition to the large-LDOS regions, we identify two more categories of regions of interest here. The first are regions which correspond to a large local $\Delta_S$. Similar to the $\omega = 0.09$ plot, the LDOS for these regions remains suppressed and is notably featureless, with no visible modulations within this regions. The second are the regions whose LDOS had peaked at energies $\omega < 0.16$. At $\omega = 0.16$, each of these regions consistently features a ring of enhanced intensity surrounding a region where the LDOS is suppressed. Finally, at $\omega = 0.31$ (Fig.~\ref{fig:ldosplots_8}d), the remaining large-LDOS regions have a smaller value of the LDOS compared to analogous regions at smaller frequencies, consistent with the observation in Fig.~\ref{fig:ldosbygap_ldoshist_8}a that the coherence-peak height is smaller for larger spectral gaps. These large-gap regions do not retain any sharp signatures of the square patches in $\Delta_T$. They appear instead as ``islands'' amid a background that hosts visible modulations. The large areas that host these modulations consist of all the regions whose local $\Delta_S$ is smaller than $\omega = 0.31$; these modulations arise from quasiparticle scattering off of the large order-parameter ``barriers'' formed by the remnant large-$\Delta_T$ regions. At this energy, these modulations are visible because the spatial extent of the all regions whose local spectral gap $\Delta_S < 0.31$ is large, on the order of several Fermi wavelengths; this is to be contrasted with the relative absence of modulations in the $\omega = 0.16$ case, for which the regions with $\Delta_S < 0.16$ are much smaller in extent.

In Fig.~\ref{fig:linecuts_8}, we show the LDOS, $\Delta_T$, $\Delta_S$, and $\Delta_L$ along a linecut through the middle of the sample. Fig.~\ref{fig:linecuts_8}a in particular shows what typical spectra look like in the system. At low energies, there is little variation in the LDOS, but at higher energies, the spectra begin to exhibit considerable inhomogeneity. The LDOS for large-$\Delta_S$ lattice sites show kinks, consistent with the bin-averaged plots in Fig.~\ref{fig:ldosbygap_ldoshist_8}a. Many of these spectra exhibit more complex features, such as multiple-peak lineshapes, that are washed out in the bin-averaged LDOS data. It can also be observed that the lineshapes evolve smoothly as a function of position, with the lineshape for a given site appearing similar to those from other sites lying within a distance $\approx \xi$ from it.

We had noted earlier that the correlation between $\Delta_T$ and $\Delta_S$ as evinced from the real-space plots in Fig.~\ref{fig:centeredplots_8} is not perfect. More pictorial evidence for this comes from comparing these two quantities along the same linecut (Fig.~\ref{fig:linecuts_8}b). $\Delta_S$ generally tracks $\Delta_T$, but we can observe that for sites where $\Delta_T$ is very small, the corresponding $\Delta_S$ is always larger and offset to the right. In contrast, for large-$\Delta_T$ sites, these two quantities match each other well. That said, $\Delta_S$ exhibits slower spatial fluctuations than $\Delta_T$ along the same linecut; rapid variations in $\Delta_T$ are not reflected by a similar amount of variation in $\Delta_S$, and these two quantities match each other best at sites lying within clusters of multiple patches with similar values of $\Delta_T$ are present. As for $\Delta_L$, we find that not only is its variation much smaller than either $\Delta_T$ or $\Delta_S$, but its variations are also much more long-ranged than the other two quantities: while $\Delta_T$ and $\Delta_S$ vary on the scale set by $l$, $\Delta_L$ on the other hand varies over longer length scales (several multiples of $l$). The sharp boundaries are largely invisible to $\Delta_L$, which only curves slightly in the presence of a boundary between small- and large-$\Delta_S$ regions. We also find that $\Delta_L$ tends to diverge from $\Delta_T$ or $\Delta_S$ most strongly when $\Delta_T$ is large; in this instance, it tends to remain near $\Delta_L \approx \overline{\Delta}$, and only increases mildly in the presence of this strong-pairing patch.

All these aforementioned observations are neatly summarized by two-dimensional histograms of the three possible pairs of order-parameter measures, which we show in Fig.~\ref{fig:histogram_opsgleg_8}. The histogram for the $\Delta_T$-$\Delta_S$ pair (Fig.~\ref{fig:histogram_opsgleg_8}a) shows that while there is an overall positive correlation between these two variables and the distributions are both broad, the correspondence retains a degree of fuzziness. A number of interesting features are present in this particular plot. First, we can see that $\Delta_S$ is nonzero throughout the system. The smallest $\Delta_S\approx 0.09$, and these small spectral gaps occur at sites where $\Delta_T \approx 0$---the vanishing of the order parameter in the $l \approx \xi$ regime never manifests itself in the concomitant vanishing of the spectral gap. At larger values of $\Delta_T$, the correspondence gradually improves to the point where $\Delta_T \approx \Delta_S$, but beyond a ``spine'' where this correspondence is particular strong, there remains a broad but faint surrounding cloud where these two values are not closely matched with one another. The $\Delta_T$-$\Delta_L$ pair (Fig.~\ref{fig:histogram_opsgleg_8}b) on the other hand shows behavior very similar to what was seen in the $l < \xi$ case (Fig.~\ref{fig:histogram_opsgleg_1}b): $\Delta_L$ is narrowly distributed, mildly fluctuates from its mean value ($0.16$), and only weakly reflects the underlying variations in $\Delta_T$, with a comparatively weak but positive correlation present between these two variables. Finally, the two-dimensional histogram of the $\Delta_S$-$\Delta_L$ pair (Fig.~\ref{fig:histogram_opsgleg_8}c) shows unusual behavior in that for small values of $\Delta_S$, much of the joint distribution of the two variables is localized within in a small region of values (like the $l < \xi$ case). However, for larger values of $\Delta_S$, their joint distribution extends out similarly to that seen in the $\Delta_T$-$\Delta_L$ pair in Fig.~\ref{fig:histogram_opsgleg_8}b, with a narrow distribution of $\Delta_L$ compared to the much broader distribution of $\Delta_S$. Evidently, these two order-parameter measures behave very differently in the $l \approx \xi$ regime, again owing to the fact that $\Delta_S$ is measured at high energies while $\Delta_L$ is measured at low energies where the system is largely homogeneous. 

We also plot the correlations of $\rho_C$ with $\Delta_T$ and $\Delta_S$ in Fig.~\ref{fig:histogram_opsgcph_8}. It can be seen that both $\Delta_T$ and $\Delta_S$ remain negatively correlated with $\rho_C$ in the $l \approx \xi$ regime, similar to the $l < \xi$ case. However, $\rho_C$ is more broadly distributed in the $l \approx \xi$ case, which is consistent with the general observation that regions with the smallest spectral gaps exhibit very sharp resonances while those with the largest $\Delta_S$ have relatively stubby coherence peaks.

In sum, what we have observed here is that for the inhomogeneous $d$-wave superconductor with $l \approx \xi$, the low-energy ($\omega < \Delta_K$) spectra are homogeneous (for the $l = 8$ results here, the low-energy spectra are that of a \emph{homogeneous} $d$-wave superconductor with $\Delta = 0.16$), while the high-energy ($\omega > \Delta_K$) LDOS exhibits much of the full inhomogeneity of the underlying order parameter, but with uneven correspondences between $\Delta_S$ and $\Delta_T$. Notably, proximity effects are strong; this is most evident in the spectra with $\Delta_T \approx 0$, which show a finite $\Delta_S$. The critical energy $\Delta_K$, which separates homogeneous and inhomogeneous states, in turn appears to be largely set by the smallest spectral gaps in the system. A crucial difference between the $l = 8$ regime studied here and the $l = 1$ one in the previous section is that the quasiparticle response of the system to order-parameter inhomogeneity largely hews to the underlying order parameter at high energies such that $\Delta_T$ and $\Delta_S$ are positively correlated, but at low energies the response reflects little of the order-parameter variation (so that $\Delta_L$ is only weakly correlated with $\Delta_T$). Evidently, increasing the inhomogeneity length scale $l$ reduces the energy range in which the spectrum is homogeneous, so that for sufficiently large $l$, the resulting critical energy scale $\Delta_K$ separating homogeneous and inhomogeneous spectra is pushed downwards towards the Fermi energy. 

\section{Large Patches ($l > \xi$)}

\begin{figure}[t]
	\centering
	\includegraphics[width=0.5\textwidth]{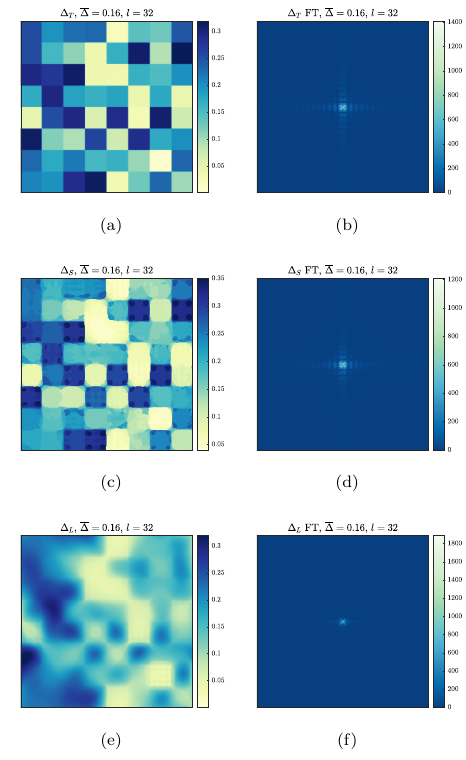}
	\caption{Plots of $\Delta_T$ (a), $\Delta_S$ (c), and $\Delta_L$ (e), and the absolute value of their respective Fourier transforms (b, d, and f) for patch size $l = 32$. Only one disorder realization is presented. Shown here are quantities taken from the middlemost $256 \times 256$ segment of the full $1024 \times 256$ system.}
	\label{fig:centeredplots_32}
\end{figure}

\begin{figure}[t]
	\centering
	\includegraphics[width=0.5\textwidth]{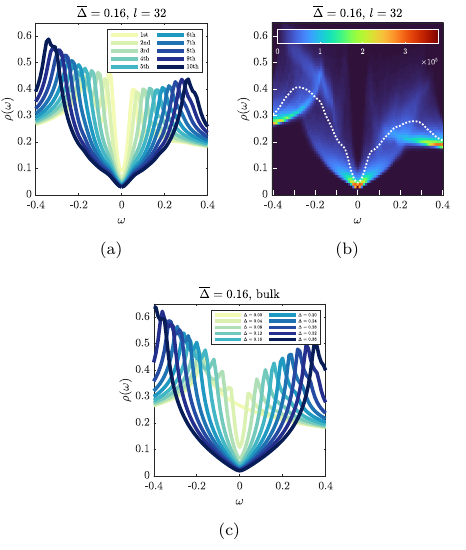}
	
	\caption{Top left: Plots of the binned average local density of states for $l = 32$. Each LDOS spectrum is binned according to its spectral gap (with 10 bins in all), and all spectra in a given bin are then averaged over. Top right: Plot of the distribution of the local density of states as a function of energy (heat map) and the average local density of states (dashed white line) for $l = 32$. The spectra are binned according to the energy, and histograms for each energy are taken. The counts per energy bin are shown as a heat map. For both these plots, 4 disorder realizations and a total of 1,048,576 individual spectra are used in this plot. Bottom: Plots of the density of states for a clean bulk $d$-wave superconductor for various values of $\Delta$. To obtain these plots, we made use of a $1024\times 1024$ momentum-space grid, and the same band-structure and broadening parameters in the inhomogeneous cases are used here.}
	\label{fig:ldosbygap_ldoshist_32}
\end{figure}

\begin{figure}[t]
	\centering
	\includegraphics[width=0.50\textwidth]{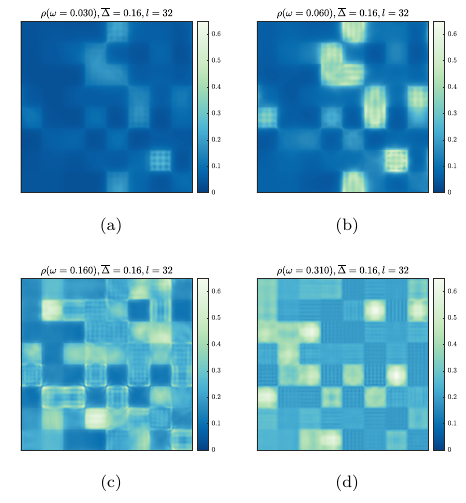}
	\caption{Plots of the LDOS at different frequencies for patch size $l = 32$. The same disorder realization and field of view as in Fig.~\ref{fig:centeredplots_32} are used here.}
	\label{fig:ldosplots_32}
\end{figure}

\begin{figure}[t]
	\centering
	\includegraphics[width=0.44\textwidth]{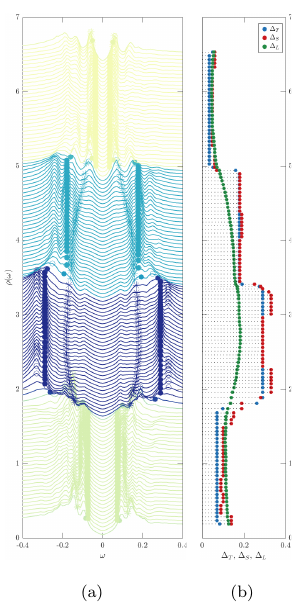}
	
	\caption{Left: Plots of the local density of states along a straight line through the middle of the sample from $(516,64)$ to $(516,192)$ for $l = 32$. The filled circles indicate the position of the underlying order parameter $\Delta_T$, while open circles and asterisks indicate the spectral gap $\Delta_S$ and low-energy gap $\Delta_L$, respectively. Only a single disorder realization is shown here. The plots are colored according to the local value of $\Delta_T$. Right: Plots of $\Delta_T$ (blue), $\Delta_S$ (red), and $\Delta_L$ (green) along the same linecut.}
	\label{fig:linecuts_32}
\end{figure}

\begin{figure}[t]
	\centering
	\includegraphics[width=0.5\textwidth]{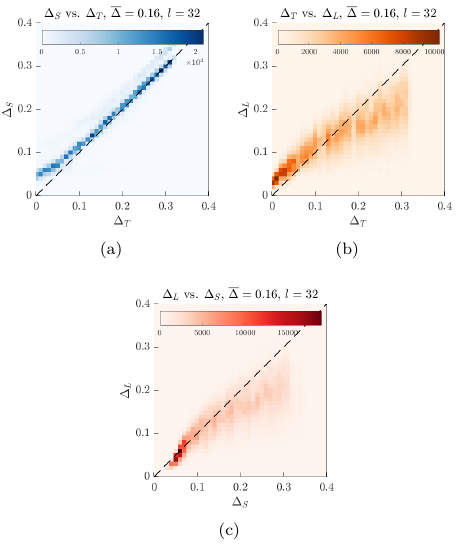}
	
	\caption{Plots of the two-dimensional histograms for a) $\Delta_T$ and $\Delta_S$; b) $\Delta_T$ and $\Delta_L$; and c) $\Delta_S$ and $\Delta_L$ for $l = 32$. Included here are data points from four disorder realizations corresponding to 1,048,576 individual LDOS spectra.}
	\label{fig:histogram_opsgleg_32}
\end{figure}

\begin{figure}[t]
	\centering
	
	\includegraphics[width=0.50\textwidth]{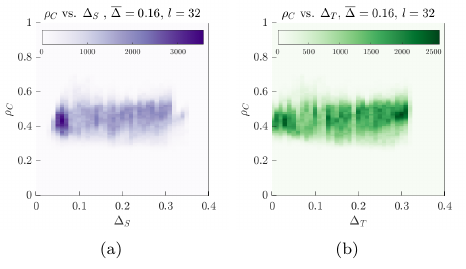}
	
	\caption{Plots of the two-dimensional histograms for a) $\Delta_S$ and $\rho_C$ and b) $\Delta_T$ and $\rho_C$ for $l = 32$. Included here are data points from four disorder realizations corresponding to 1,048,576 individual LDOS spectra.}
	\label{fig:histogram_opsgcph_32}
\end{figure}

To complete the list of examples, we finally consider the case where large patches with $l > \xi$ are present. The bulk homogeneous $d$-wave superconductor can be thought of as the  $l \to \infty$ limit of this model. A particular interest of ours in this regime is the extent to which the behavior expected from a clean superconductor already makes itself manifest, and the extent to which the residual ``granularity'' of the system (in the form of boundaries between different patches)  continues to affect various observables. As before, we set $\overline{\Delta} \approx 0.16$, and the patches are of size $l = 32$ so that we are solidly in the $l > \xi$ regime. Maps of $\Delta_T$, $\Delta_S$, and $\Delta_L$ (both real-space and Fourier-transformed) from a $256 \times 256$ section of the $l = 32$ system are shown in Fig.~\ref{fig:centeredplots_32}.

In this regime, both $\Delta_T$ and $\Delta_S$ maps look mostly identical to one another; this can be seen in  Fig.~\ref{fig:centeredplots_32}a and c. The difference is primarily in the more rounded shape of the patches in $\Delta_S$, and in the fact that  when adjacent patches in $\Delta_T$ have similar (but not identical) values, they show up in $\Delta_S$ as a large, mostly uniform patch, with the gradations in $\Delta_T$ not visible in $\Delta_S$. Also, isolated small-$\Delta_S$ regions tend to be slightly larger in spatial extent than similarly isolated large-$\Delta_S$ regions, an observation that is also clearly seen in the $l \approx \xi$ case; this is a reflection of the larger local coherence length within these small-gap patches, which ensures that (through the proximity effect) the effect of the small-$\Delta_T$ patch ``leaks out'' into the surrounding large-$\Delta_T$ region. Conversely, a large-$\Delta_T$ patch surrounded by a small-$\Delta_T$ would feature the ``leakage'' of the small-gap background into the patch, making the large-$\Delta_S$ patch look smaller than its actual size in $\Delta_T$ is. This effect for $l > \xi$ however is much more muted compared to what is seen in the $l \approx \xi$ case. Finally, similar to the $l = 1$ and $l = 8$ cases discussed earlier, the $\Delta_S$ map always is nonzero, with a minimum $\Delta_S$ present even in regions where $\Delta_T$ is close to zero. Even with $l = 32$, proximity effects are strong enough that a nonzero pair density, and consequently a finite gap, is present deep within these patches. This minimum gap however is smaller than that seen in the $l \approx \xi$ case.

The map of the low-energy gap $\Delta_L$ with $l  = 32$ (Fig.~\ref{fig:centeredplots_32}e) now clearly reflects the inhomogeneous nature of the underlying order parameter. Small- and large-$\Delta_T$ regions match up neatly with small- and large-$\Delta_L$ regions, and this time the distribution of $\Delta_T$ is broad enough to capture the variations in $\Delta_T$ to reasonable accuracy. The agreement is clearly not perfect, though: the finer gradations seen in $\Delta_T$ are not captured by $\Delta_L$, and the distribution of values of $\Delta_L$ still appears to be narrower than that of $\Delta_S$. Nevertheless, the $\Delta_L$ plots demonstrate that for $l > \xi$, the low-energy states now display more inhomogeneity than those in the $l \approx \xi$ and $l  < \xi$ cases, but still not as much as the high-energy states do. The overall similarity among the $\Delta_T$, $\Delta_S$, and $\Delta_L$ maps carries over to their respective Fourier transforms (Figs.~\ref{fig:centeredplots_32}b, d and f). In particular, the Fourier transforms of the $\Delta_T$ and $\Delta_S$ maps look almost identical, with both featuring a highly localized region of high intensity near zero wavevector and higher harmonics arising from the regular square shape of the patches in both real-space maps. The Fourier transform of the $\Delta_L$ map also features strong intensities only at small wavevectors. One can see that the small-momentum part of Fig.~\ref{fig:centeredplots_32}f is similar to that of Figs.~\ref{fig:centeredplots_1}f  and~\ref{fig:centeredplots_8}f, but as higher-momenta scattering processes are suppressed for the $l = 32$ case, the only feature in the Fourier transform left is a much smaller X-shaped feature near zero momentum.

These results lend further credence to the supposition mentioned in the previous section that the energy scale $\Delta_K$ gets pushed down in energy with increasing inhomogeneity length scale $l$. The bin-averaged LDOS (Fig.~\ref{fig:ldosbygap_ldoshist_32}a) shows that while the low-energy part of the spectra remain more homogeneous than the high-energy part, the low-energy spectra become sufficiently differentiated from one another in that the small-gap spectra are vertically offset relative to the large-gap ones by a finite amount, and that the low-energy slopes of the various bin-averaged LDOS spectra vary more strongly; this differentiation among the spectra is more prominent compared to the $l = 1$ and $l = 8$ cases considered earlier. This is the reason why the $\Delta_L$ maps for $l  = 32$ look much more inhomogeneous than in the other cases. Kinks in the spectra remain, however, and like in the $l = 8$ case, the characteristic energy $\Delta_K$ where these kinks appear seems to be set once more by the smallest spectral gaps in the system (witness how the location of the kink in the largest-$\Delta_S$ spectrum coincides more or less with the smallest binned $\Delta_S$). $\Delta_K$ for $l = 32$ is smaller than that in the $l = 1$ and $l = 8$ cases. Note that while the spectra shown in Fig.~\ref{fig:ldosbygap_ldoshist_32}a look more bulk-like, many features of the bulk case (Fig.~\ref{fig:ldosbygap_ldoshist_32}c) still do not manifest fully. In particular, for $l = 32$, the spectral gap of the bin-averaged spectra appears to be only weakly correlated with the coherence-peak height, whereas in the bulk case, these two quantities are strongly \emph{positively} correlated with one another. We note as well the wide distributions of the LDOS as shown in Fig.~\ref{fig:ldosbygap_ldoshist_32}b, which gives rise to an \emph{average} LDOS with an unusual lineshape, with a V-shaped low-energy part and a broad shoulder-like feature at higher energies. In the presence of this type of inhomogeneity, it is then very difficult to tell from the average LDOS lineshape alone what the true average value of $\Delta_T$ is. Naively taking the maximum of the average LDOS will give a huge overestimate of $\overline{\Delta}$. 

Real-space plots of the LDOS (Fig.~\ref{fig:ldosplots_32}) meanwhile show behavior consistent with Fig.~\ref{fig:ldosbygap_ldoshist_32}. As with the two cases ($l = 1$ and $l = 8$) looked at in detail earlier, we consider four representative frequencies, namely $\omega = 0.03$, $\omega = 0.06$, $\omega = 0.16$, and $\omega = 0.31$. The low-energy case ($\omega = 0.03$, Fig.~\ref{fig:ldosplots_32}a) is mostly homogeneous, but compared to plots from the $l = 1$ and $l = 8$ cases taken at the same frequency, the underlying patchwork order-parameter inhomogeneity is more visible: the small-$\Delta_T$ patches already show a larger LDOS compared to the background. Interestingly, the smallest-$\Delta_S$ regions already exhibit modulations at these energies; evidently, the large extent of the patches means that quasiparticle scattering within the ``wells'' formed by the surrounding large-$\Delta_T$ barriers occurs even at frequencies smaller than the local spectral gap within these patches. When the frequency is increased to $\omega = 0.06$ (Fig.~\ref{fig:ldosplots_32}b), the small-gap regions now display a very large LDOS. As with the $l = 8$ case, no modulations can be seen in the large-$\Delta_T$ regions. 

Similar to the $l = 8$ case, at $\omega = 0.16$ (Fig.~\ref{fig:ldosplots_32}c), there are three kinds of regions: large-LDOS patches which correspond to patches where the local spectral gap $\Delta_S \approx 0.16$; small-LDOS patches which correspond to large-$\Delta_S$ regions with $\Delta_S > 0.16$; and regions hosting prominent modulations which correspond to small-$\Delta_S$ regions with $\Delta_S < 0.16$. These striking modulations are (like those seen in the $l = 8$ case) a result of quasiparticle scattering off of the walls of the order-parameter barrier formed by the large-$\Delta_T$ neighboring patches. These regions with modulations grow in size with increasing $\omega$. At $\omega = 0.31$ (Fig.~\ref{fig:ldosplots_32}d), the only regions without any visible modulations are the large-LDOS patches corresponding to patches with large $\Delta_S$; these modulations are present everywhere else. A peculiarity of our square-patch disorder model is that when $l$ is large (such as in the present $l = 32$ case), the boundaries of the small-$\Delta_S$ regions that host these modulations are straight and have right-angle corners, which result in mostly ordered modulations in the $(1,0)$ and $(1,0)$ directions. This is in contrast with that seen in the $l = 8$ case earlier where, as a result of the small value of $l$, the regions that have these modulations are irregularly shaped and therefore the resulting scattering patterns are not as sharply defined as in the present case.

The LDOS along a cut through the sample (Fig.~\ref{fig:linecuts_32}a) shows that the LDOS lineshapes within a single patch are largely uniform, except for sites near boundaries between patches. Like in the $l = 8$ case, various unusual features emerge in the lineshapes taken in the vicinity of these boundaries, such as the presence of two peaks and in-gap kinks. Finally, plots of $\Delta_T$ and $\Delta_S$ along this same cut (Fig.~\ref{fig:linecuts_32}b) show that these two quantities follow each other very closely; the spectral gap $\Delta_S$ provides an almost faithful representation of the underlying order parameter $\Delta_T$, even for small $\Delta_T$. $\Delta_L$ on the other hand still displays a marked tendency to deviate from either $\Delta_S$ or $\Delta_T$ at strong-pairing regions; it tends to agree with these other two quantities when $\Delta_T$ is small or intermediate (i.e., $\Delta_T \approx \overline{\Delta}$) in size. Its variations are also much more long-wavelength than those of $\Delta_S$ or $\Delta_T$, which vary over the intrinsic length scale of the underlying inhomogeneity; $\Delta_L$, like in the $l = 8$ case, appears to be largely blind to the sharp boundaries between patches (unlike $\Delta_S$) and instead varies smoothly as one crosses patch boundaries.

To succinctly summarize all these observations together, we show once more the various two-dimensional histograms of the order-parameter measures in Fig.~\ref{fig:histogram_opsgleg_32}. The corresponding plot for the $\Delta_T$-$\Delta_S$ pair (Fig.~\ref{fig:histogram_opsgleg_32}a) shows that the two quantities are now very strongly correlated with one another, with only minimal fuzziness in the correspondences. It is worth noting that the system still exhibits a finite $\Delta_S$ everywhere, even in regions where $\Delta_T \approx 0$. Here, the smallest $\Delta_S \approx 0.05$, which is smaller than the smallest $\Delta_S$ for the $l \approx \xi$ case. This is consistent with the observation made earlier that the kink energy $\Delta_K$, which is the scale that separates homogeneous and inhomogeneous states and which looks to be tied to the smallest spectral gaps in the system, is pushed down in energy as the inhomogeneity length scale $l$ is increased. The histogram for the $\Delta_T$-$\Delta_L$ pair (Fig.~\ref{fig:histogram_opsgleg_32}b) now shows a much more pronounced positive correlation between these two quantities compared with the two cases considered before, with the agreement best at small $\Delta_T$ and worst at large $\Delta_T$. Finally, the $\Delta_S$-$\Delta_L$ pair (Fig.~\ref{fig:histogram_opsgleg_32}c) also appears to be more correlated with one another than in the previous cases. As with the $l \approx \xi$ case, here we find a similar tendency for the joint distribution to be mostly localized within one small region for small $\Delta_S$ and for it to extend out for increasing $\Delta_S$. 

In Fig.~\ref{fig:histogram_opsgcph_32}, we show the two-dimensional histograms of $\rho_C$ taken with $\Delta_T$ or $\Delta_S$. We can see in either plot that $\rho_C$ shows almost no correlations with $\Delta_T$ or $\Delta_S$: $\rho_C$ does not vary with either order-parameter measure. This is consistent with Fig.~\ref{fig:ldosbygap_ldoshist_32}a, where it can be seen that the coherence peaks of the bin-averaged spectra do not vary appreciably with increasing spectral gap. Evidently, even with $l = 32$, we clearly are not yet within the fully bulk regime (see Fig.~\ref{fig:ldosbygap_ldoshist_32}c) where $\rho_C$ is strongly positively correlated with $\Delta_T$ or $\Delta_S$.

From what we have seen in the $l > \xi$ example shown here, it is apparent that bulk-like behavior is approached in this regime; however, various proximity-coupling effects arising from the boundaries between patches result in unusual spectra that still display the characteristics of inhomogeneous superconductivity at smaller inhomogeneity length scales, such as kinks in the spectrum and residual, albeit much weaker, low-energy homogeneity. However, we can generically expect these effects to become even weaker with increasing $l$. As $l$ is made larger, the volume fraction occupied by the boundaries between the patches becomes smaller, so these effects associated with a rapidly changing order parameter contribute less and less to the LDOS and to the statistics of various quantities, allowing one to recover bulk behavior in the $l \to \infty$ limit.

\section{Low-Energy Homogeneity and Spectral Kinks}

\begin{figure}[t]
	\centering
	
	\includegraphics[width=0.5\textwidth]{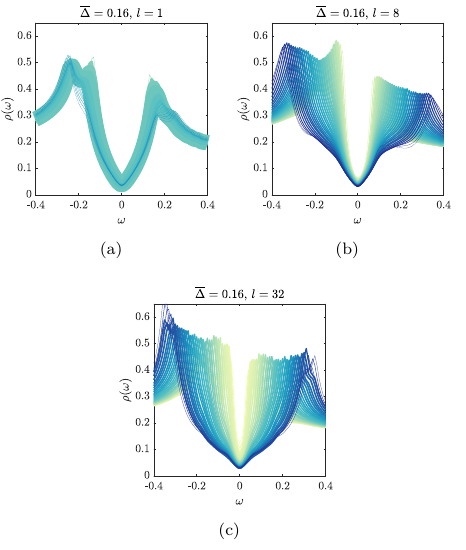}

	\caption{Plots of the binned average local density of states for $l = 1$ (a), $l = 8$ (b), and $l = 32$ (c), with finer bins used than in the previous plots. Each LDOS spectrum is binned according to its spectral gap (with 41 bins in all), and all spectra in a given bin are then averaged over. For these plots, 4 disorder realizations and a total of 1,048,576 individual spectra are used. The width of each line plot is proportional to the number of spectra within the bin. Only lineshapes corresponding to bins with total number of spectra greater than 1024 are shown here.}
	\label{fig:ldosbygap_fine_016}
\end{figure}

\begin{figure}[t]
	\centering

	\includegraphics[width=0.5\textwidth]{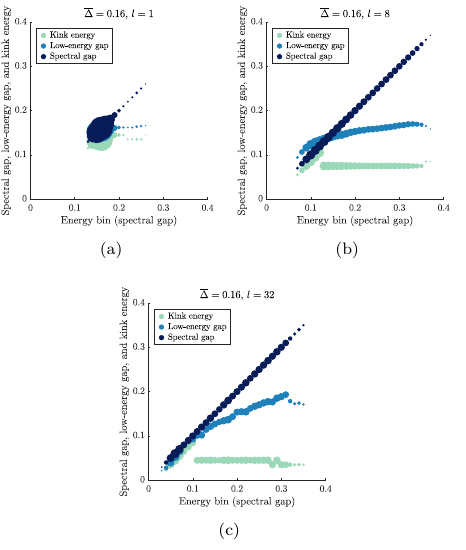}
	
	\caption{Plots of the spectral gap $\Delta_S$, the low-energy gap $\Delta_L$, and the kink energy $\Delta_K$ for a) $l = 1$, b) $l = 8$, and c) $l = 32$ for $\overline{\Delta} = 0.16$. The data presented here is derived from Fig.~\ref{fig:ldosbygap_fine_016}. The size of each marker is proportional to the number of spectra within the bin.}
	\label{fig:gapcomparison_1_8_32}
\end{figure}

\begin{figure}[t]
	\centering
	\includegraphics[width=0.5\textwidth]{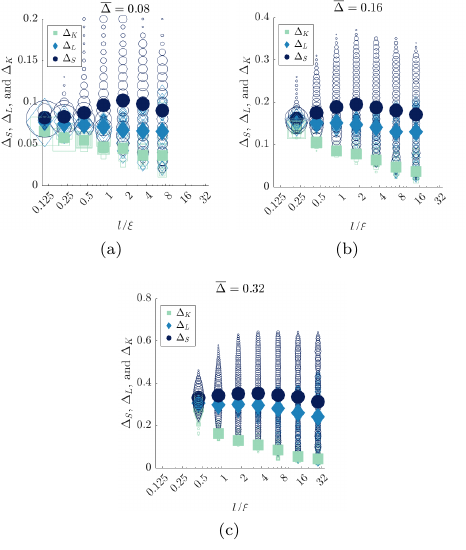}
	
	\caption{Plots of the spectral gap $\Delta_S$, the low-energy gap $\Delta_L$, and the kink energy $\Delta_K$ of the bin-averaged spectra for a) $\overline{\Delta} = 0.08$, b) $\overline{\Delta} = 0.16$, and c) $\overline{\Delta} = 0.32$ as a function of $l/\xi$. The open markers indicate values of the relevant quantities corresponding to each bin-averaged spectrum, while the filled markers show the average of the said quantity over all spectra. In Fig.~\ref{fig:gapcomparison_allgaps}b ($\overline{\Delta} = 0.16$), these quantities are extracted from the bin-averaged spectra shown in part in Fig.~\ref{fig:ldosbygap_fine_016}. The same $x$-axis scale is used in all three plots. The size of each marker is proportional to the number of spectra within the bin.}
	\label{fig:gapcomparison_allgaps}
\end{figure}

Perhaps the most notable behavior seen in STS experiments on the cuprates is the sharp dichotomy between the low-energy homogeneous states and the high-energy inhomogeneous states that are characteristic of the $l \gtrapprox \xi$ case, which is accompanied by the presence of kinks appearing in large-$\Delta_S$ spectra. We had noticed in Figs.~\ref{fig:ldosbygap_ldoshist_8}a and~\ref{fig:ldosbygap_ldoshist_32}a that these kinks tend to appear at a characteristic kink energy $\Delta_K$ which is roughly set by the smallest spectral gaps of the system. In this section, we define $\Delta_K$ appropriately and study it as a function of $l$, contrast its behavior with the other energy scales that can be obtained from the LDOS spectra ($\Delta_S$ and $\Delta_L$), and put on quantitative footing the informal observations made in the previous section.

We first define $\Delta_K$ in quantitative terms. One can observe in Figs.~\ref{fig:linecuts_8}a and~\ref{fig:linecuts_32}a that a purely \emph{local} definition of $\Delta_K$ as the energy where a kink is present is difficult to implement in practice in our numerics: at low energies, numerous wiggles in the LDOS as a function of energy can be seen, and it is difficult to ascertain if these are the true kinks that separate the homogeneous states from the inhomogeneous ones, or if these arise from finite-size effects. This difficulty of isolating the kink in the spectrum is otherwise not present in our operational definitions of $\Delta_S$ (which involves simply finding the peak of the positive-energy LDOS) and $\Delta_L$ (which involves taking the slope of the LDOS near $\omega = 0$).

To get around this difficulty, we work instead with the bin-averaged spectra, as visualized in Fig~\ref{fig:ldosbygap_fine_016}. Here, we bin the individual spectra by its local spectral gap. Unlike Figs.~\ref{fig:ldosbygap_ldoshist_1}a, ~\ref{fig:ldosbygap_ldoshist_8}a,  and~\ref{fig:ldosbygap_ldoshist_32}a, however, here we use much smaller bin widths that allow us to capture the averaged spectrum for each value of $\omega$. Because of the very large number of individual spectra, even a finer binning procedure allows for enough spectra per bin to ensure that an average over a sufficiently large number of spectra is performed for each individual bin. The averaging over a large number of spectra allows us to smoothen out any finite-size effects, and we can thus define $\Delta_K$, as well as $\Delta_S$ and $\Delta_L$, for each of the bin-averaged spectra. $\Delta_K$ is defined as the lowest energy at which the \emph{derivative} with respect to $\omega$ of the positive-energy bin-averaged LDOS exhibits a peak. Note that this definition implies that $\Delta_K \approx \Delta_S$ \emph{in the absence of a kink}. Right away, we can expect the following behaviors to hold at these two extreme limits: a) in the extreme pointlike limit $l \ll \xi$ where the states are highly homogeneous, all three energy scales coincide, i.e., $\Delta_S \approx \Delta_L \approx \Delta_K \approx \text{constant}$; and b) in the extreme bulk limit $l \to \infty$, $\Delta_S \approx \Delta_T$ while $\Delta_K \to 0$. 

In Fig.~\ref{fig:gapcomparison_1_8_32}, we show the extracted $\Delta_S$, $\Delta_L$, and $\Delta_K$ for the $l = 1$, $l = 8$, and $l = 32$ cases (with $\overline{\Delta} = 0.16$), taken from the data presented in Fig.~\ref{fig:ldosbygap_fine_016}. One can see that in the $l = 1$ case (Fig.~\ref{fig:gapcomparison_1_8_32}a, corresponding to $l < \xi$), all three energy scales largely coincide with one another and only weakly vary, which is a reflection of the homogeneity of the LDOS at all energies. The fact that $\Delta_K \approx \Delta_S$ meanwhile is due to the absence of kinks in most of the spectra. For the $l = 8$ case (Fig.~\ref{fig:gapcomparison_1_8_32}b), we now observe that these three energy scales cleanly separate from one another. $\Delta_S$ varies most strongly, while $\Delta_L$ does so only weakly, consistent with Fig.~\ref{fig:histogram_opsgleg_8}. The kink energy $\Delta_K$ is a constant, except for bins corresponding to the smallest spectral gaps (for which \emph{no} kink is found, and therefore its $\Delta_K$ value is determined by its $\Delta_S$ value). It can be seen that where the kink is a well-defined feature in the averaged spectra, $\Delta_K$ is determined primarily by the value of $\Delta_S$ of the lowest bins, since these averaged spectra do not contain kinks. Finally, for the $l = 32$ case (Fig.~\ref{fig:gapcomparison_1_8_32}c), the $\Delta_L$ line now has a steeper slope, reflecting a broader distribution of $\Delta_L$; its slope is still less than that of the $\Delta_S$ line, however. $\Delta_K$ again is largely constant for the bin-averaged spectra that show kinks, and is determined by $\Delta_S$ of the first bin, as in the $l = 8$ case. $\Delta_K$ is smaller in value for $l  = 32$ compared to its value for $l = 8$, entirely in line with the observations made in the previous section that the energy scale at which inhomogeneity sets in becomes lower as $l$ is increased. These are consistent with the expectations we had outlined earlier for the extreme limits $l \ll \xi$ and $l \to \infty$. In particular, in going from $l = 8$ to $l = 32$, we can see that $\Delta_L \to \Delta_S$ and $\Delta \to 0$. In the physically interesting $l \approx \xi$ regime (Fig.~\ref{fig:gapcomparison_1_8_32}b), however, it is clear that these three energy scales appear to show the most divergent behavior from one another.

In Fig.~\ref{fig:gapcomparison_allgaps}, we show these three quantities together on a single semi-log plot as a function of $l/\xi$ for three values of the average order parameter ($\overline{\Delta} = 0.08$, $\overline{\Delta} = 0.16$, and $\overline{\Delta} = 0.32$, corresponding to average coherence lengths of $\xi \approx 9.06$, $\xi \approx 4.53$, and $\xi \approx 2.27$, respectively). In these plots, for each value of $l$, we have plotted each of the bin-averaged values of $\Delta_S$, $\Delta_L$, and $\Delta_K$ together, along with the \emph{average} of all bin-averaged values weighted by the number of spectra in each bin (shown as filled markers). (Showing the average of the bin-averaged spectra, rather than the corresponding quantity computed from the system- and disorder-averaged spectra, allows us to sidestep any ambiguities in defining the spectral gap $\Delta_S$ from taking system-wide averages---compare Figs.~\ref{fig:ldosbygap_ldoshist_32}a and b to see an example of how the system- and disorder-averaged LDOS spectrum can yield a misleading average $\Delta_S$.) For all three $\overline{\Delta}$ values considered, the behaviors of the three energy scales are the same. At the smallest values of $l$, all three energy scales cluster together closely. As $l$ increases, however, these three scales start to diverge from one another. The average $\Delta_S$ increases with increasing $l$, and appears to peak within the $l \approx \xi$ regime before starting to decrease. Meanwhile, the average $\Delta_L$ decreases as $l$ is increased, albeit only very weakly, not straying far from $\overline{\Delta}$. Finally, $\Delta_K$ can be seen to decrease monotonically with increasing $l$. A close examination of $\Delta_K$ shows that it generally tracks very closely with the values of $\Delta_S$ of the lowest bins. This result shows that the appearance of these kinks coincides with the very strong resonances in the LDOS that appear when the frequency $\omega$ equals the smallest spectral gaps in the system.

These results suggest that in the physically relevant $l \approx \xi$ regime, the notion of an average spectral gap---defined either by taking both the system- and configuration-average of $\Delta_S$ or by binning the LDOS and obtaining the bin-averaged $\Delta_S$, as we have done here---leads to a considerable overestimate of the true average underlying order parameter $\overline{\Delta}$. This is partly due to the finite $\Delta_S$ even in regions where $\Delta_T \approx 0$: these finite $\Delta_S$ values push up the average, leading to an overestimate of $\overline{\Delta}$. As we had noted earlier, these particular $\Delta_S$ values (i.e., the smallest gaps) are correlated with the value of $\Delta_K$; therefore, as $l$ increases, $\Delta_K$, and therefore these particular $\Delta_S$'s, decrease toward zero, and the average of $\Delta_S$ regresses to $\overline{\Delta}$ for sufficiently large $l$. It is telling that in Fig.~\ref{fig:gapcomparison_allgaps}, even patches as large as $l = 64$ for the $\overline{\Delta} = 0.32$ case still feature a large enough discrepancy between the averages of $\Delta_S$ and $\Delta_L$ to indicate that the system in this case is still far from the extreme bulk limit. On the other hand, taking the average of $\Delta_L$ appears to give a more accurate estimate of $\overline{\Delta}$, owing to the robustness of low-energy homogeneity when $l \approx \xi$.

\section{$l$-Dependence of Correlations Between Order-Parameter Measures}

\begin{figure}[t]
	\centering
	\includegraphics[width=0.5\textwidth]{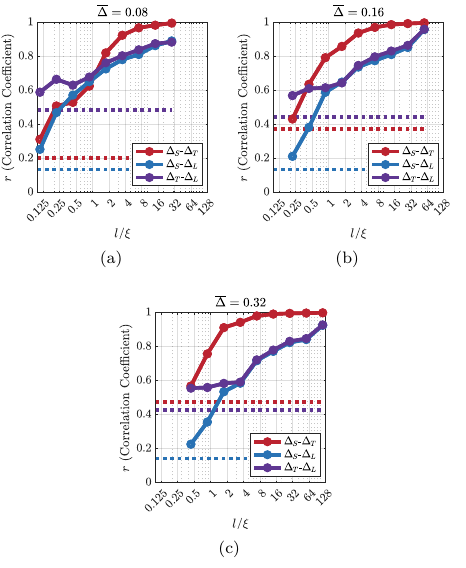}
	
	\caption{Plots of the correlation coefficient $r$ between $\Delta_T$ and $\Delta_S$ (orange), $\Delta_S$ and $\Delta_L$ (blue), and $\Delta_T$ and $\Delta_L$ (purple) as a function of patch size $l$ for various values of the average order parameter. The dashed lines indicate the values of $r$ for the ''$l = 0$'' case where the bond order parameters are independently and identically randomly distributed. The same $x$-axis scale is used in all three plots.}
	\label{fig:correlation_opsgleg}
\end{figure}

\begin{figure}[t]
	\centering
	\includegraphics[width=0.5\textwidth]{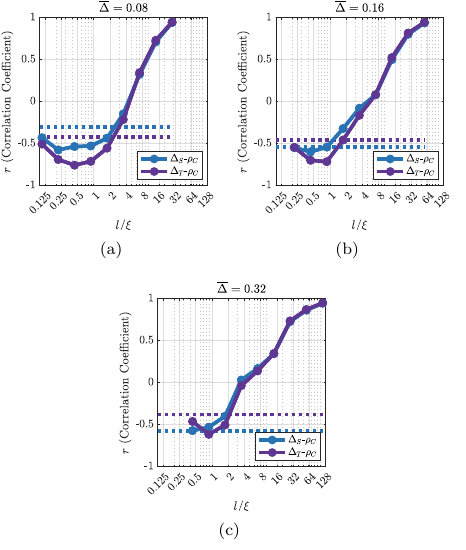}
	
	\caption{Plots of the correlation coefficient $r$ between $\Delta_S$ and $\rho_C$ (blue), and between $\Delta_T$ and $\rho_C$ (purple) as a function of patch size $l$ for various values of the average order parameter. The dashed lines indicate the values of $r$ for the ''$l = 0$'' case where the bond order parameters are independently and identically randomly distributed. The same $x$-axis scale is used in all three plots.}
	\label{fig:correlation_opsgcph}
\end{figure}

In this section, we fully quantify the observations made in the preceding sections about the correlations present among various order-parameter measures ($\Delta_T$, $\Delta_S$, and $\Delta_L$) and the coherence-peak height ($\rho_C$) and examine these as a function of the inhomogeneity length scale $l$. To assess the correlation between a pair of variables, we use the Pearson correlation coefficient $r$: a perfectly linear correlation between two variables gives $r = 1$, the absence of any correlation gives $r = 0$, and a perfectly linear anticorrelation results in $r = -1$. As such, this quantity is an excellent measure of the accuracy of the order-parameter measurables ($\Delta_S$ and $\Delta_L$) as compared with the true underlying order parameter ($\Delta_T$). To obtain $r$, we collect all values of the relevant pair of quantities measured on all sites and across four disorder realizations, giving us $1,048,576$ data points per quantity.

In Fig.~\ref{fig:correlation_opsgleg}, we plot $r$ as a function of $l$ for the following pairs of quantities: $\Delta_S$-$\Delta_T$, $\Delta_S$-$\Delta_L$, and $\Delta_T$-$\Delta_L$. We consider once more three values of the mean order parameter $\overline{\Delta}$: $\overline{\Delta} = 0.08$, $\overline{\Delta} = 0.16$, and $\overline{\Delta} = 0.32$. For the $\Delta_S$-$\Delta_T$ pair, the behavior of $r$ is to a good extent independent of $\overline{\Delta}$. The correlation can be seen to be very small when $l \ll \xi$. At the smallest value of $l/\xi$ considered ($l = 1$ for $\overline{\Delta} = 0.08$, corresponding to $l \approx 0.11\xi$), $r$ for this pair of variables is only around 0.3. (This correlation is actually even smaller for the maximally disordered ``$l = 0$'' case where the bond order parameters are independently distributed, for which the correlation drops to around 0.2.) When $l < \xi$, the coefficient generally is less than 0.65, but in the physically interesting regime where $l \approx \xi$, $r$ falls within the range $r \in [0.6, 0.9]$, with the correlation being weaker for $\overline{\Delta} = 0.08$ and stronger for $\overline{\Delta} = 0.32$. When $l \gtrapprox 4\xi$, the correlation between this pair of variables is very strong, with $r > 0.9$. That the correspondence between $\Delta_S$ and $\Delta_T$ improves with increasing $l/\xi$ is not surprising, but it is reassuring that in the regime relevant to the cuprates ($l \approx \xi$), $\Delta_S$ is a fairly accurate, if imperfect, proxy for the true order parameter when order-parameter inhomogeneity is present.

Meanwhile, the correspondence between $\Delta_L$ and $\Delta_T$ is considerably weaker compared with that between the $\Delta_S$-$\Delta_T$ pair, except when $l/\xi < 2$ for $\overline{\Delta} = 0.08$ and $l/\xi < 0.5$ for $\overline{\Delta} = 0.16$. While $r$ increases monotonically with $l$ in all cases, we find that $r$ at large values of $l / \xi$ (such as $l / \xi \approx 8$) is still below the value of $0.9$ seen in the $\Delta_S$-$\Delta_T$ case (at which point $r$ will have saturated to almost unity in that case). The difference between $r$ for the $\Delta_L$-$\Delta_T$ pair and that for the $\Delta_S$-$\Delta_T$ pair is most pronounced in the $\overline{\Delta} = 0.32$ case. We find additionally that $r$ generally depends on $l$, but \emph{not} on $l / \xi$: regardless of the value of $\overline{\Delta}$ used, the obtained $r$ remains largely the same for fixed $l$. When $l  = 1$, $r \approx 0.6$, and this value barely changes until $l = 8$ is reached, at which point $r$ increases. But even for $l = 32$, $r \approx 0.8$, suggesting an imperfect correspondence even at very large patch sizes. This is presumably due to the lingering presence of kinks even at large patch sizes which continue to homogenize the slope of the low-energy density of states, making $\Delta_L$ more uniform than the underlying order parameter $\Delta_T$ even within this regime. Similar statements can be made about the correlation of the $\Delta_S$-$\Delta_L$ pair: $r$ in this case is consistently smaller than that of the $\Delta_S$-$\Delta_T$ case (except for a narrow window of $l$ values in the $\overline{\Delta} = 0.08$ case). Also, $r$ appears to depend on $l$ alone and not on $l /\xi$. Interestingly, $r$ for $\Delta_S$-$\Delta_L$ and $r$ for $\Delta_T$-$\Delta_L$ differ from one another at small values of $l$, but start to match almost exactly when $l \geq 8$. The reason for this is the fact that both $\Delta_S$ and $\Delta_L$ are very narrowly distributed for small $l$ and their joint distribution consequently appears as a small, highly localized ellipse (see Fig.~\ref{fig:histogram_opsgleg_1}c for an illustration), leading to a small $r$ in this regime.

The correlation coefficients for $\Delta_S$-$\rho_C$ and $\Delta_T$-$\rho_C$ pairs behave similarly as a function of $l/\xi$ for all values of $\overline{\Delta}$ considered (see Fig.~\ref{fig:correlation_opsgcph}). When $l/\xi < 4$, $r$ for both pairs is negative, and only when $l/\xi$ is increased beyond this threshold does the correlation become positive. In the $l \approx \xi$ regime, $r$ for the $\Delta_S$-$\rho_C$ pair generally lies between $-0.3$ and $-0.6$, while $r$ for $\Delta_T$-$\rho_C$ falls within the $[-0.4,-0.7]$ range. Strong correlations ($r \approx 0.7$ and above) between $\rho_C$ and either order-parameter measure only start to appear at very large patch sizes with $l / \xi \gtrapprox 32$; the fact that the correspondence remains imperfect even with large patches suggests that the bulk regime is still not fully realized, and that boundary effects between patches play a role in spoiling the expected strong correlation between the coherence-peak height and the order parameter in the large-patch regime.

\section{Application to Experiments}

We have studied in detail the role of the length scale of order-parameter inhomogeneity on various observables, using a simple patch model which manages to reproduce (for $l \approx \xi$) many features seen in the $dI/dV$ spectra of the cuprates across various dopings such as low-energy homogeneity, spectral kinks, and the negative correlation between the spectral gap $\Delta_S$ and the coherence-peak height $\rho_C$. In this section, we discuss what our results mean in the context of various experiments performed on the cuprates.

\subsection{The Width of the Distribution of the Order-Parameter Inhomogeneity}

In our simulations, we have found that when the patch sizes are smaller than or on the order of the average coherence length $\xi$, the distribution of the spectral gap $\Delta_S$ is smaller than that of the underlying order parameter $\Delta_T$. In the case where $l < \xi$, the spectral gap distribution is extremely narrow, but even when $l \approx \xi$, we see that regions where $\Delta_T$ is almost zero nevertheless feature finite values of $\Delta_S$. As we had noted earlier, this is due to proximity coupling between the patches. The pair density ``leaks out'' from the larger-$\Delta_T$ patch to the smaller one, extending out over a length scale $\approx \xi$; because the patches themselves are of linear dimension $\approx \xi$, the pair density never goes to zero throughout the patch.

These results open up the possibility that the cuprates are even more inhomogeneous than presently supposed from estimates taken from the width of the spectral-gap distribution. Across a broad doping range (from $p = 0.08$ to $p = 0.22$), it is found that the standard deviation $\sigma$ of $\Delta_S$ for Bi-2212 is $\sigma \approx 0.2 \overline{\Delta_S}$ \cite{mcelroy2005coincidence,alldredge2008evolution,gomes2007visualizing,pasupathy2008electronic}. If we suppose that the underlying order-parameter ($\Delta_T$) inhomogeneity in the cuprates is in the $l \approx \xi$ regime, then $\sigma$ is an \emph{underestimate} of the width of the $\Delta_T$ distribution. In particular, it is conceivable that there exist patches where the pairing interaction vanishes but the local spectral gap is nonzero due to the proximity effect. 

No direct comparison of the distributions of $\Delta_S$ and $\Delta_T$ has been performed thus far for the cuprates, although it is now to obtain a direct image of the underlying order parameter via Josephson STS methods, which use superconducting flakes as tips in the scanning tunneling microscope \cite{hamidian2016detection,cho2019strongly,du2020imaging}.  A local measurement of the critical current $I_c$ using these Josephson tunnel-junction techniques turns out to be a faithful representation of $\Delta_T$ \cite{graham2017imaging,graham2019josephson,sulangi2021correlations}. That said, such a comparison has already been made in an iron-based superconductor (FeTe$_{0.55}$Se$_{0.55}$), where it was found that the $I_c$ distribution (and consequently that of $\Delta_T$) is considerably broader than that of the spectral gap \cite{cho2019strongly}. While there are differences between that system and the cuprates (e.g., the pairing symmetry of FeSeTe is different from that of the cuprates, and stronger chemical-potential disorder is present in FeSeTe), this particular material provides a concrete example of a system for which the width of the spectral-gap distribution severely underestimates that of the order-parameter distribution.

In the absence of a comparison between experimentally obtained $\Delta_S$ and $\Delta_T$, the correlations we obtained involving the coherence-peak height can be used as a guide to estimate the length scale of the underlying order-parameter inhomogeneity. STS experiments generally find that $\Delta_S$ and $\rho_C$ are negatively correlated with one another across all dopings where superconductivity is present. In our simulations, we find this to be true when $l < 4\xi$, regardless of the value of $\overline{\Delta}$ used; this provides an upper bound for the length scale of the $\Delta_T$ inhomogeneity. This is further evidence for the $l \approx \xi$ scenario that is believed to be the case in the cuprates across a broad doping range. 

We finally note that in the $l \approx \xi$ regime, the real-space map of $\Delta_S$ (Fig.~\ref{fig:centeredplots_8}b) looks very similar to the various gap maps from various STS experiments in the cuprates (e.g., Fig. 1 of Fang et al. \cite{fang2006gap}, Fig. 1 of McElroy et al. \cite{mcelroy2005coincidence}, and Fig. 5a of Ye et al. \cite{ye2024emergent}), the square shape of the underlying patches notwithstanding. This demonstrates that the underlying order parameter can, at least in principle, look very different from the spectral gap map. In our model, we assumed that the patches in $\Delta_T$ are all of the same size, yet the resulting $\Delta_S$ map shows that the regions with the smallest gaps have the largest spatial extent and vice versa, similar to what is seen in STS (e.g., Fang et al. \cite{fang2006gap}). Our results suggest the possibility that the small- and large-$\Delta_T$ regions all have the same typical size: the $\Delta_S$ measurements provide an overestimate and an underestimate of the spatial extent of the small- and large-$\Delta_T$ patches, respectively.

\subsection{Origin of Kinks and Low-Energy Homogeneity}

\begin{figure}[t]
	\centering
	\includegraphics[width=0.5\textwidth]{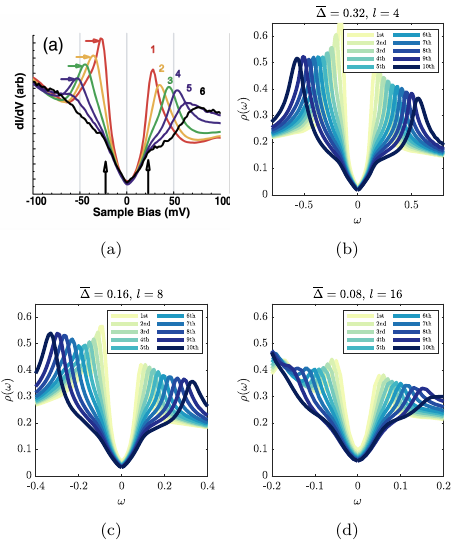}
	
	\caption{Top left: Plots of the binned average differential conductance $dI/dV$ taken from an underdoped ($T_c = 79$ K, $p \approx 15$\%) Bi-2212 sample. Reprinted figure with permission from McElroy \emph{et al.}, \emph{Phys. Rev. Lett} \textbf{94} 197005 (2005) \cite{mcelroy2005coincidence}. Copyright (2005) by the American Physical Society. Top right, bottom left, and bottom right: Plots of the binned average local density of states taken from simulated data with $\Delta = 0.32$, $l = 4$ (b); $\Delta = 0.16$, $l = 8$ (c); and $\Delta = 0.08$, $l = 16$ (d), respectively. All three of these scenarios fall within the $l \approx \xi$ regime.}
	\label{fig:ldosbygap_theory_vs_experiment}
\end{figure}

Consistently seen in STS studies of many cuprates are kinks in the differential conductance spectra, which appear at a characteristic energy $\Delta_K$ that is parametrically less than $\Delta_S$. In underdoped-to-optimally doped Bi-2212, these kinks are ubiquitous and prominent, and found to be present mainly in large-gap regions \cite{mcelroy2005coincidence,alldredge2008evolution,pushp2009extending,alldredge2013universal}. For overdoped Bi-2212, these are also present in the spectra, albeit in more muted form, with the kink energy closer to the average spectral gap \cite{pushp2009extending,alldredge2013universal}. Meanwhile, in Bi-2201, these kinks remain prominent seen even at moderate overdoping, close to the end of the superconducting dome \cite{ye2024emergent}. The origin of these kinks is not fully known, and it is generally believed that since these kinks are more prominent at small dopings, these are an effect of strong correlations. However, our work provides a clear picture of their origin from a mean-field perspective: these arise from proximity coupling between patches with different order-parameter values. This proximity effect is strongest at low energies, where the quasiparticle response is very similar to that of a homogeneous $d$-wave superconductor with order parameter $\overline{\Delta}$. This homogeneity is lost past the kink energy $\Delta_K$, which serves as a dividing line between the homogeneous states and the inhomogeneous ones. The kinks therefore arise from the system finally starting to reflect the underlying inhomogeneity for $\omega > \Delta_K$ and deviating from the homogeneous lineshapes at low-energies. This also explains why the coherence-peak heights are inversely correlated with the spectral gap: much of the spectral weight that would have been in the large-gap coherence peak in the absence of inhomogeneity is put instead in the kink, resulting in a shorter coherence peak. Note that this mechanism does \emph{not} invoke a frequency-independent self-energy that arises separately from inhomogeneity (e.g., from inelastic scattering). We return to this point later.

This mesoscopic superconducting proximity-effect-based mechanism appears to explain well the kinks seen in both Bi-2212 and Bi-2201 across a broad doping range from underdoping to overdoping. In particular, Fig.~\ref{fig:ldosbygap_ldoshist_8}a resembles very closely Fig.~2a of McElroy et al. \cite{mcelroy2005coincidence}, which describes bin-averaged spectra taken from a mildly underdoped ($T_c = 79$ K) Bi-2212 sample. We provide a side-by-side comparison of the McElroy et al. results with a more extensive selection of our simulated data (with $\overline{\Delta} = 0.08$, $\overline{\Delta} = 0.16$, and $\overline{\Delta} = 0.32$)  in Fig.~\ref{fig:ldosbygap_theory_vs_experiment}. Fig.~\ref{fig:ldosbygap_theory_vs_experiment}d also looks very similar to Figs.~1e and 1f of Ye et al. \cite{ye2024emergent}, which describes binned LDOS spectra for optimally doped and mildly overdoped Bi-2201. (One can observe from Fig.~\ref{fig:ldosbygap_theory_vs_experiment}d that the kinks are less prominent the smaller $\overline{\Delta}$ is, a likely explanation for their decreasing prominence in the overdoped regime.) For these two particular experimental datasets, the kink energies appear to coincide with the smallest spectral gap, which supports the proximity-effect mechanism outlined earlier. 

However, at \emph{strong} underdoping, the mechanism outlined here is arguably too simplistic to account for the particularly severe kinks seen in this regime. In particular, Pushp et al. \cite{pushp2009extending} and McElroy et al. \cite{mcelroy2005coincidence} both find that \emph{nearly all} spectra in this regime show kinks. Within our model, some of the spectra (the small-gap ones) should not exhibit kinks at all, so this effect at strong underdoping appears to lie beyond our mechanism. This likely signals a breakdown of the mean-field $d$-wave description in this regime. We hypothesize that the kinks at severe underdoping arise not only from the superconducting proximity effect, but also from strong correlation effects resulting from proximity to the nearby antiferromagnetic insulating phase, which together have the effect of ``pushing out'' the spectral gaps as frequency is increased, resulting in most spectra having kinks as opposed to only the large-gap ones.

\subsection{Near-Node Gap Measurements}

\begin{figure}[t]
	\centering
	\includegraphics[width=0.5\textwidth]{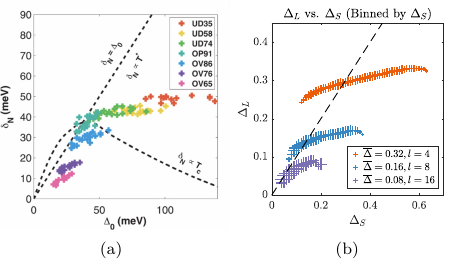}

	\caption{Left: Plot of the extracted inverse nodal slope ($\delta_N$, $y$-axis) and the antinodal gap ($\Delta_0$, $x$-axis) from $dI/dV$ spectra taken from Bi-2212 at various dopings. From Ref.~\cite{pushp2009extending}, Pushp \emph{et al}. Reprinted with permission from AAAS. Right: Plots of the \emph{binned} $\Delta_L$ vs. $\Delta_S$ with $\Delta = 0.08$, $l = 16$ (purple); $\Delta = 0.16$, $l = 8$ (blue); and $\Delta = 0.32$, $l = 4$ (orange), taken from simulated data. The binning is done by $\Delta_S$, and corresponds to data previously shown in Fig.~\ref{fig:gapcomparison_allgaps}. The dashed line indicates $\Delta_T$ = $\Delta_S$. The size of each marker is proportional to the number of spectra within the corresponding bin. All three of these scenarios fall within the $l \approx \xi$ regime.}
	\label{fig:histogram_opsgleg_theory_vs_experiment}
\end{figure}

Pushp et al. perform a systematic doping-dependent analysis and find that the underdoped cuprates exhibit a ``universal'' low-energy regime where ``nodal gap''---i.e., an estimate of the order parameter derived from either the slope of the local density of states near $\omega = 0$ (which is equivalent to our definition of the low-energy gap $\Delta_L$) or from the result of their fitting procedure taken at the gap node---is strikingly doping-independent from the strongly underdoped regime all the way to optimal doping \cite{pushp2009extending}. This universality is lost beyond optimal doping: for overdoped cuprates, they find that nodal gap decreases with increasing doping. They also find that as doping is lowered, the kink energies become smaller and that $\Delta_L$ is less than all binned $\Delta_S$ for heavily underdoped Bi-2212 ($T_c = 35$ K and $T_c = 58$ K). It should be noted that in this regime, the large-$\Delta_S$ spectra show extremely rounded coherence peaks that do not resemble those arising from a mean-field model of $d$-wave superconductivity; this is also seen in the STS data from McElroy et al. taken in this regime \cite{mcelroy2005coincidence}.

Both sets of observations (underdoped and overdoped) can be explained, at least to a reasonable extent, within the inhomogeneous $d$-wave model. For overdoped Bi-2212, it appears that the average $\overline{\Delta}$ of the order parameter $\Delta_T$ decreases as hole doping increases. This is consistent with the picture that, as doping is increased, correlation effects become less important and the relevant physics reverts to Fermi-liquid theory in the normal state and BCS theory in the superconducting state, with a weaker pairing interaction the farther one gets from the strongly correlated parent state.

Moderately underdoped and optimally doped Bi-2212 on the other hand appears to be well described by a \emph{doping-independent} $\overline{\Delta}$, which is necessary to ensure the observed behavior of $\Delta_L$. The fact that the kink energy as a fraction of the average spectral gap increases with increasing doping is consistent with the decrease of the inhomogeneity length scale $l$ (with fixed $\xi$, since $\overline{\Delta}$ is roughly constant) and/or the decrease of the width of the inhomogeneity distribution as doping is increased. It should be noted that McElroy et al. \cite{mcelroy2005coincidence}, Gomes et al. \cite{gomes2007visualizing}, Alldredge et al. \cite{alldredge2008evolution}, and Pushp et al. \cite{pushp2009extending} find that the width of the $\Delta_S$ distribution decreases with increasing doping. Both possibilities (decreasing $l$ or width of the $\Delta_T$ distribution) are consistent with this observation. Recall that in Fig.~\ref{fig:gapcomparison_allgaps}, we see that even though the underlying order-parameter distributions have the same mean value $\overline{\Delta}$, the resulting average of the spectral gaps can nevertheless vary as a function of $l$, with the maximum difference between $\overline{\Delta}$ and the average $\Delta_S$ occuring when $l /\xi \approx 2$-$4$. Taking this into consideration, the average value of $\Delta_S$ as inferred from experiment is likely to be an \emph{overestimate} of $\overline{\Delta}$ in the cuprates. The observed increase of the average $\Delta_S$ and the accompanying decrease in $\Delta_K$ with decreasing doping could likely arise from an increase in the intrinsic inhomogeneity length scale ratio $l$. Much of the doping-dependent phenomenology of $\Delta_S$, $\Delta_L$ and $\Delta_K$ described in Ref.~\cite{pushp2009extending}, in Fig. 6b of Ref.~\cite{alldredge2008evolution}, and in Fig. 13b in Ref.~\cite{alldredge2013universal} can thus be captured by the plots in Fig.~\ref{fig:gapcomparison_allgaps} which show what happens to these various energy scales when both $\overline{\Delta}$ and the inhomogeneity width are both kept fixed while $l$ increases (keeping $\xi$ fixed). Having said this, we note that these behaviors could just as well be described by a decrease in the width of the inhomogeneity distribution with increasing doping: the kink energy is determined by the smallest spectral gaps which would increase when the width of the inhomogeneity is decreased while holding $\overline{\Delta}$ constant. This is corroborated in Appendix A.

We note finally that Fig.~\ref{fig:histogram_opsgleg_8}c, which describes the joint distribution of $\Delta_S$ and $\Delta_L$ for the $l \approx \xi$ case, is very similar to Fig. 4a of Pushp et al. \cite{pushp2009extending} (which describes the same quantities, but using bin-averages) for a variety of dopings ranging from moderate underdoping ($T_c = 58$ K) to moderate overdoping ($T_c = 65$ K). (An explicit comparison of the experimental data from Pushp et al. with our $l \approx \xi$ simulated results with $\overline{\Delta} = 0.08$, $\overline{\Delta} = 0.16$, and $\overline{\Delta} = 0.32$, using bin-averaged data to replicate fully the presentation of the Pushp et al. results, is shown in Fig.~\ref{fig:histogram_opsgleg_theory_vs_experiment}.) Both the experimental and simulated data show similar features: the $\Delta_L$ distributions are much narrower than those of $\Delta_S$, and, rather strikingly, the joint distribution mainly extends out towards the $\Delta_L < \Delta_S$ portion of the plot from $\Delta_L \approx \Delta_S$, with comparatively little weight in the $\Delta _L > \Delta_S$ portion of the plot. Within our model, this form of the joint distribution is due to the fact that a relatively large minimum spectral gap is present even in regions where the order parameter is zero, so much of the weight has piled onto the smallest $\Delta_S$'s (where this weight would have been more spread out in the case of the joint distribution of $\Delta_T$ and $\Delta_L$). That said, we note that in Fig.~\ref{fig:histogram_opsgleg_theory_vs_experiment}, the experimental joint distributions of $\Delta_S$ and $\Delta_L$ are  such that these lie entirely within the $\Delta_L < \Delta_S$ portion of the plot. We are not sure as to why this is the case. In our numerical work, in the $l \approx \xi$ regime, we generally find that the average spectral gap $\Delta_S$ is generally larger than the average low-energy gap $\Delta_L$, but nevertheless a small but nonzero fraction of spectra satisfy $\Delta_L > \Delta_S$. As far as the cuprates are concerned, this shift may signal the presence of subtle self-energy effects not included in our analysis that renormalize the spectral gap upward at higher energies, but should these be the reason underlying this effect, it is remarkable that that these persist even at overdoping, where correlation effects are expected to be weaker. It should be said that this shift is less pronounced when $\overline{\Delta}$ is small, so correlation effects may not be needed to be invoked in the overdoped case; see the plot for $\overline{\Delta} = 0.08$ in Fig.~\ref{fig:histogram_opsgleg_theory_vs_experiment}, which shows that nearly all of the weight of the joint distribution lies \emph{below} the $\Delta_L = \Delta_S$ line.
 
\subsection{Self-Energy Fits}

Many analyses of STS data on the cuprates employ a procedure in which the individual $dI/dV$ spectra are fitted to a ``Dynes-like'' formula involving a $d$-wave order parameter, a suitable normal-state tight-binding model, and a scattering rate $\Gamma$ to account for the broadening of spectra. Various models of scattering rates have been used in the literature. Commonly used is an energy-independent but spatially varying scattering rate \cite{pasupathy2008electronic,tromp2023puddle}, which ensures that the broadening is uniform at all energies. More sophisticated approaches include an energy- and position-dependent scattering rate (e.g., $\Gamma(\mathbf{r}_i,\omega) \propto \omega$, as used in Refs.~\cite{alldredge2008evolution} and~\cite{alldredge2013universal}) which takes into account inelastic scattering from electron-electron interactions. Such an energy-dependent model is able to account on a phenomenological level for the shorter and broadened coherence peaks in large-gap spectra taken from underdoped cuprates.

It is interesting to revisit these scattering-rate estimates in light of our studies of $d$-wave superconductors with inhomogeneous pairing in the $l \approx \xi$ regime. We generally find that without the explicit inclusion of an energy-dependent scattering rate (recall that we use an energy-independent $\Gamma$), the coherence peaks of large-$\Delta_S$ regions are already shorter than expected from a clean bulk $d$-wave superconductor with the same $\Delta_S$ and $\Gamma$. These suggest that the scattering-rate estimates made in Refs.~\cite{alldredge2008evolution} and~\cite{alldredge2013universal} are an \emph{overestimate}, as their fits do not take into account the inhomogeneity of the order parameter. The fitting procedure they employ assumes that each local spectrum is captured by a clean model of a $d$-wave superconductor with some $\Delta_T(\mathbf{r}_i)$ and $\Gamma(\mathbf{r}_i,\omega)$. However, we have already seen from our examples earlier in this paper that the identification of the local $\Delta_S$ and the true local $\Delta_T$ is not exact, especially in the $l \approx \xi$ case relevant to Bi-2212 and Bi-2201. As mentioned earlier, the origin of the shorter coherence peaks in the $l \approx \xi$ regime are the low-energy homogeneity and the kinks: as the low-energy spectra are homogeneous, the large-gap spectra feature much more spectral weight at low energies compared with the \emph{bulk $d$-wave superconductor with the same large spectral gap}, leaving less spectral weight for the high-energy coherence peak and rendering it comparatively shorter. While the large-gap spectra do feature considerable broadening to suggest that inelastic scattering plays an important role (especially in heavily underdoped cuprates), the amount of inelastic scattering needed to reproduce the spectra appears to be less than originally estimated.

The scattering-rate estimates made in Refs.~\cite{alldredge2008evolution} and~\cite{alldredge2013universal} can be reinterpreted in light of our numerics in the following way. These quantify the amount of spectral weight that is transferred away from the coherence peaks to other parts of the spectrum, regardless of the mechanism for the spectral-weight transfer (inhomogeneity, inelastic scattering, etc.). Order-parameter inhomogeneity transfers spectral weight to the kink, while inelastic scattering broadens out the coherence peak directly and redistributes spectral weight mainly in the vicinity of the peak. These scattering-rate estimates do not distinguish between these two contributions to spectral-weight transfer. In principle, it would be possible to disentangle these two effects by numerically computing the spectra of the inhomogeneous system and iteratively adjusting $\Delta_T(\mathbf{r}_i)$ and $\Gamma(\mathbf{r}_i,\omega)$ until decent fits to the experimental spectra are obtained.

\subsection{Extinction of Quasiparticle Scattering Interference}

Quasiparticle scattering interference is a ubiquitous phenomenon in the cuprates; however, it is famously known to disappear in the underdoped cuprates when the bias voltage is raised above a critical threshold. This phenomenon (dubbed ``QPI extinction'') has been observed in underdoped and optimally doped Bi-2212 \cite{kohsaka2008cooper,fujita2014simultaneous} and Bi-2201 \cite{he2014fermi,webb2019density}, but \emph{not} at overdoping, where QPI continues to be present at all energies below the mean spectral gap of the system. Strikingly, Kohsaka et al. find that the energy at which QPI disappears is the \emph{same} energy at which the homogeneity of the low-energy states is lost \cite{kohsaka2008cooper}. This last observation provides an important clue to pinpointing the true origin of QPI extinction.

Our results on inhomogeneous $d$-wave superconductors provide a possible reason for this phenomenon. At low energies below $\Delta_K$, the quasiparticle states are homogeneous, so coherent scattering can occur when these quasiparticles encounter short-ranged quenched disorder. In this regime, the QPI response should be that of a uniform $d$-wave superconductor with order parameter $\overline{\Delta}$. Above $\Delta_K$, on the other hand, the states are now inhomogeneous: the LDOS now reflects the local value of $\Delta_T$, and sharp differences between the LDOS on nearby sites are present. We had seen in our discussion of Figs.~\ref{fig:ldosplots_8} that for energies $\omega > \Delta_K$, the system divides itself into three classes of regions corresponding to whether the local spectral gap $\Delta_S$ at any particular site satisfies $\omega \approx \Delta_S$, $\omega > \Delta_S$, or $\omega < \Delta_S$, with the quasiparticle states localized within any one of these three classes of regions. Thus, the coherent propagation of wavelike quasiparticle excitations across the entire system ceases to be possible when $\omega > \Delta_K$, and therefore QPI is no longer present. Details of this argument, making use of a model of inhomogeneous $d$-wave superconductivity with weak scatterers present, will be presented in a future paper \cite{sulangifuture2}.

While compelling, a crucial weakness of this picture is that it does not explain why $\Delta_K$ (i.e., the energy where both low-energy homogeneity and QPI disappear) exactly coincides with the energy where the tips of the contours of constant energy (CCE) of the Bogoliubov quasiparticles crosses the antiferromagnetic zone boundary. Within our model of inhomogeneous $d$-wave superconductivity, there is no reason why $\Delta_K$ should match this energy, which would be set by the band-structure parameters and $\overline{\Delta}$. This additional requirement likely signals that additional ingredients beyond an inhomogeneous $d$-wave order parameter should be present; this is likely where the broken-symmetry states at high energies play a key role. Other proposals for resolving the problem of QPI extinction have been put forth, invoking coexisting antiferromagnetic order \cite{andersen2009extinction} and modulated hoppings near impurities \cite{vishik2009momentum}; these proposals explain why QPI disappears when the tips of the CCEs extend past the antiferromagnetic zone boundary, but not why this energy coincides with the loss of homogeneity. A full explanation of this phenomenon will likely have to involve a combination of inhomogeneity and some of these other ingredients, in addition to squaring fully with the highly correlated nature of the underdoped superconducting state.

\section{Conclusion}

In this paper, we have extensively studied the effect of changing the length scale of order-parameter inhomogeneity. We find that within the simple model of broadly distributed and patchy ($l \approx \xi$) inhomogeneity, we are able to reproduce much of the striking phenomenology seen in STS experiments on the cuprates: the dichotomy between the low-energy (homogeneous) and high-energy (inhomogeneous) regimes; spectral kinks; and the ``universality'' of the ``nodal gap.'' The remarkable feature of these results is that these arise without needing to invoke any additional coexisting broken-symmetry order. In particular, the kinks, whose origins have long been a puzzle in the cuprate lore, are now seen to have a particularly simple explanation: these are due to the superconducting proximity effect, which tends to homogenize the states at low energies, but whose effect becomes lost at the kink-energy scale $\Delta_K$. 

Our results will hopefully spur a reexamination of the proper interpretation of kinks and the low-energy--high-energy dichotomy seen in many STS experiments on the cuprates. It has long been thought that the kink energy is the superconducting energy scale, while the spectral-gap energy is the pseudogap scale brought about by the presence of coexisting order. We find that both of these energy scales can have the \emph{same} origin---a broadly distributed and spatially extended inhomogeneous $d$-wave order parameter. If, as we strongly suspect, both of these energy scales in the cuprates originate from the same source, then many questions naturally arise for future experimental and theoretical work in this area centered on the interplay between the inhomogeneous $d$-wave superconductivity in these materials and the presence of coexisting order. In particular, it is worth asking what role coexisting order plays in the cuprates when a surprisingly large extent of the experimental phenomenology at low temperatures is adequately accounted for by gap inhomogeneity. More importantly, some of the discrepancies between our theoretical mean-field models and experiment are surely due to beyond-mean-field effects (e.g., correlation effects, superconducting fluctuations) or coexisting order not taken into account in our simulations. Our work provides a building block with which one can tease out the beyond-mean-field ingredients needed to fully characterize cuprate phenomenology.

\begin{acknowledgments}
	
It is a pleasure to thank S.-D. Chen, A. Kreisel, M. Pal, and especially W. A. Atkinson, P. J. Hirschfeld, and the late J. Zaanen for very inspiring discussions, collaborations, and encouragement, and J. C. S. Davis for graciously giving permission to reprint Fig.~\ref{fig:ldosbygap_theory_vs_experiment}a. M.A.S. acknowledges the hospitality of the Stanford Institute for Materials and Energy Sciences, where this work was initiated, and UFIT Research Computing for providing computational resources and support that have contributed to the research results reported in this publication. 
 
\end{acknowledgments}

\section{Data Availability}

The data that support the findings of this article are openly available \cite{sulangi_2025_15854143}. 

\appendix
\section{Effects of Adjusting the Width of the Order-Parameter Distribution }

\begin{figure*}[t]
	\centering
	\includegraphics[height=0.3\textwidth]{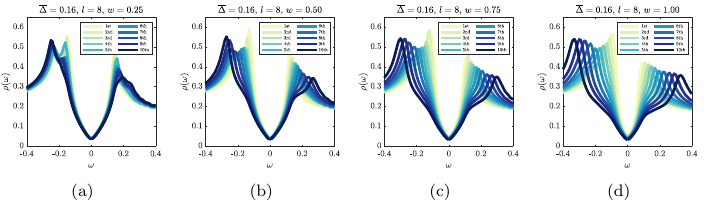}
	
	\caption{Plots of the binned average local density of states for $\overline{\Delta} = 0.16$ and $l = 8$, and different values of the width parameter $w$. Each LDOS spectrum is binned according to its spectral gap (with 10 bins in all), and all spectra in a given bin are then averaged over.}
	\label{fig:ldosbygap_width}
\end{figure*}

\begin{figure}[t]
	\centering
	\includegraphics[height=0.4\textwidth]{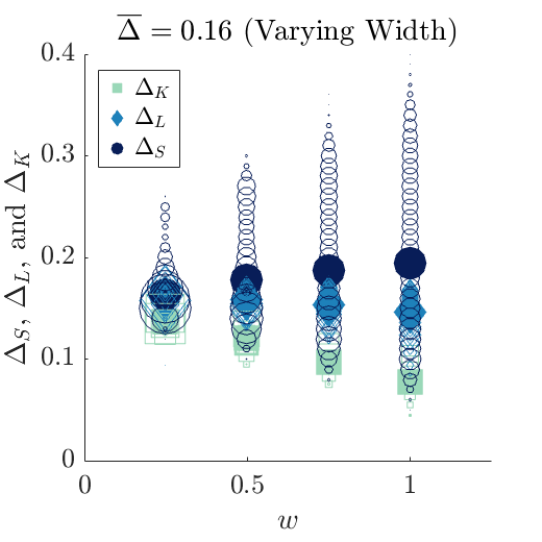}
	
	\caption{Plots of the spectral gap $\Delta_S$, the low-energy gap $\Delta_L$, and the kink energy $\Delta_K$ of the bin-averaged spectra for  $\overline{\Delta} = 0.16$, $l = 8$ as a function of $w$. The open markers indicate values of the relevant quantities corresponding to each bin-averaged spectrum, while the filled markers show the average of the said quantity over all spectra. The size of each marker is proportional to the number of spectra within the bin.}
	\label{fig:gapcomparison_width}
\end{figure}

\begin{figure*}[t]
	\centering
	\includegraphics[height=0.9\textwidth]{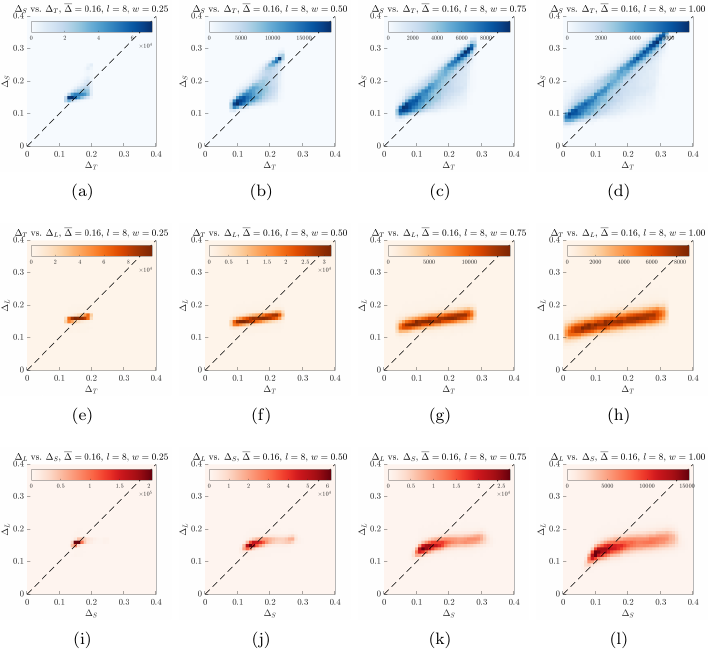}
	
	\caption{Plots of the two-dimensional histograms for $\Delta_T$ and $\Delta_S$ ((a) to (d)); $\Delta_T$ and $\Delta_L$ ((e) to (h)); and $\Delta_S$ and $\Delta_L$ ((i) to (l)) for $\overline{\Delta} = 0.16$ and $l = 8$, and different values of the width parameter $w$. The same inhomogeneity configurations are used for different values of $w$; the widths of these distributions are rescaled by $w$.}
	\label{fig:2dhist_width}
\end{figure*}

In this section, we clarify the role of the width of the order-parameter distribution on the various observables of interest to us. To this end, we make use of the $\overline{\Delta} = 0.16$, $l = 8$ model (with $l \approx \xi$) first shown in Sec. IV. We use the same inhomogeneity configurations as  presented in that section, but this time, we adjust the width of the distribution by a scaling factor $w$ so that the resulting width is $2w\overline{\Delta}$; the mean is fixed at $\overline{\Delta}$. (For instance, if $w = 0.50$, then the values that $\Delta_{D,i}$ can take lie in the range $[0.5\overline{\Delta}, 1.5\overline{\Delta}]$.) As we have already considered the maximally inhomogeneous cases in the paper ($w = 1$), we restrict ourselves to $w < 1$ and consider the following values of $w$: $0.25$, $0.50$, and $0.75$. (Cases where $w > 1$ are indeed possible, but these necessarily feature order-parameter sign reversals, which introduce attendant new physics of their own, and we neglect these in our present paper.)

In Fig.~\ref{fig:ldosbygap_width}, we plot the bin-averaged LDOS for different values of $w$. It can be seen that in the narrowest case ($w = 0.25$), the various spectra are very homogeneous and exhibit only minimal variations; the inhomogeneity only becomes apparent at energies near $\omega \approx \overline{\Delta}$, and even then, the resulting gaps vary only very mildly. This is very similar to the results we obtained for the maximally inhomogeneous small-patch model ($l < \xi$, $w = 1$) we studied in detail in Sec. III. As $w$ is increased, the kink energy becomes smaller, while the distribution of gaps becomes broader. One can see that regardless of the value of $w$ used, the kink energy tends to coincide with the smallest gaps in the system. (These observations are more succinctly plotted in Fig.~\ref{fig:gapcomparison_width}, which shows bin-averaged quantities such as the kink energy and the spectral gap as a function of $w$.) In the maximally inhomogeneous ($w = 1$) case, the smallest gaps are found in patches where $\Delta_T \approx 0$; in the non-maximally inhomogeneous cases, these smallest gaps are instead found in regions where $\Delta_T$ is nonzero.  This makes it clear that the peculiarities of the maximally inhomogeneous model, with regions where the superconducting order parameter vanishes, are not essential to the physics we have highlighted in this paper. All that is needed is that the distribution of the order parameter be sufficiently broad to see the effects of interest to experimental studies of the cuprates.

Fig.~\ref{fig:2dhist_width} shows two-dimensional histograms featuring pairs of order-parameter measures for varying $w$. A noteworthy feature of the plots for all three pairs of measures is that $w < 1$ histograms look simply like ``squished'' versions of the $w = 1$ plots. This is most evident for the histograms for $\Delta_T$-$\Delta_S$: the overall profiles of the joint distributions look similar across $w$. It is interesting to compare the $w = 0.25$ case (Fig.~\ref{fig:2dhist_width}a) with the $l < \xi$ case described in Sec. III (Fig.~\ref{fig:histogram_opsgleg_1}a): the $\Delta_S$ distribution appears similar, even though the $\Delta_T$ distribution is much narrower for $w = 0.25$ and $l = 8$ than it is for $w = 1$ and $l = 1$. These results highlight the role that the inhomogeneity length scale plays in the resulting variation in the spectral gap $\Delta_S$: inhomogeneity with small length scales effectively results in a homogeneous quasiparticle response even when the order parameter is broadly distributed. Put differently: for a fixed width of the spectral-gap distribution, the width of the underlying order-parameter distribution has to be greater (smaller) if a smaller (larger) $l$ is used.

These results allow us to interpret previous findings in a new light. The ``nematic glass'' scenario studied by Lee et al. \cite{lee2016cold} features spatially extended order-parameter inhomogeneity with $l \approx \xi$, but with a low-energy--high-energy dichotomy which does not feature kinks at a parametrically lower energy than than the average spectral gap. Their results actually resemble ours for very narrow distributions of spatially extended order-parameter inhomogeneity (e.g., Fig.~\ref{fig:ldosbygap_width}a) where the energy at which homogeneity is lost coincides with the (very narrowly distributed) spectral gap. Because the inhomogeneous order parameter in Lee et al. is obtained by self-consistency with a spatially \emph{uniform} pairing interaction, it might be the case that the resulting distribution is narrow, hence the absence of kinks in their spectra.


\bibliography{paper_ldos_inhomogeneity}

\end{document}